\documentclass[journal]{IEEEtran}
\usepackage{amsmath,amsfonts}
\usepackage{algorithmic}
\usepackage{algorithm}
\usepackage{array}
\usepackage[caption=false,font=large,labelfont=rm,textfont=rm]{subfig}
\usepackage{textcomp}
\usepackage{stfloats}
\usepackage{url}
\usepackage{verbatim}
\usepackage{graphicx}
\usepackage{cite}
\usepackage{amssymb}
\usepackage{hyperref}
\usepackage{multirow}
\usepackage{siunitx}
\usepackage{array} 
\newtheorem{property}{Property}
\newtheorem{myDef}{Definition}
\newtheorem{corollary}{Corollary}
\newtheorem{remark}{Remark}

\newtheorem{example}{Example}
\newtheorem{proof}{\it Proof}
\newcommand*{\QEDA}{\hfill\ensuremath{\blacksquare}}

\hypersetup{
    colorlinks=true,
    linkcolor=blue,
    urlcolor=blue,
    citecolor=blue
}

\begin{document}
\title{Extremely Large-scale Array Systems: Near-Field Codebook Design and Performance Analysis}
\author{Feng Zheng, Hongkang Yu, Chenchen Wang, Luyang Sun, Qingqing Wu and Yijian Chen \thanks{Feng Zheng, Chenchen Wang, Luyang Sun are with the School of Information and Communication Engineering, Beijing University of Posts and Telecommunications, Beijing 100876, China. (e-mails: zhengfeng@bupt.edu.cn, cchhen59@163.com, sly1105@bupt.edu.cn;). 

Hongkang Yu and Yijian Chen are with the State Key Laboratory of Mobile Network and Mobile Multimedia Technology, Shenzhen 518055, China. They are also with Wireless Research Institute, ZTE Corporation, Shenzhen 518055, China. (email: chen.yijian, yu.hongkang@zte.com.cn)

Qingqing Wu is with the Department of Electronic Engineering, Shanghai Jiao Tong University, Shanghai 200240, China(e-mail: qingqingwu@sjtu.edu.cn). 
}}
\date{March 2023}
\maketitle
\begin{abstract}
Extremely Large-scale Array (ELAA) promises to deliver ultra-high data rates with increased antenna elements. However, increasing antenna elements leads to a wider realm of near-field, which challenges the traditional design of codebooks. In this paper, we propose novel near-field codebook schemes based on the fitting formula of codewords' quantization performance. First, we analyze the quantization performance properties of uniform linear array (ULA) and uniform planar array (UPA) codewords. Our findings reveal an intriguing property: the correlation formula for ULA codewords can be represented by the elliptic formula, while the correlation formula for UPA codewords can be approximated using the ellipsoid formula. Building on this insight, we propose a ULA uniform codebook that maximizes the minimum correlation based on the derived formula.
Moreover, we introduce a ULA dislocation codebook to further reduce quantization overhead. Continuing our exploration, we propose UPA uniform and dislocation codebook schemes. Our investigation demonstrates that oversampling in the angular domain offers distinct advantages, achieving heightened accuracy while minimizing overhead in quantifying near-field channels. Numerical results demonstrate the appealing advantages of the proposed codebook over existing methods in decreasing quantization overhead and increasing quantization accuracy.
\end{abstract}


\section{Introduction}
\label{sec:1}
Massive multiple-input multiple-output (MIMO) technology is vital to fifth-generation (5G) mobile communication networks. Massive MIMO involves the utilization of multiple antennas to concentrate signal power within a limited area, contributing to enhanced energy efficiency and spectral efficiency \cite{ref1,ref2}. However, with the explosive demand increase on data rates in the forthcoming sixth-generation (6G) mobile communication networks, massive MIMO cannot meet the requirement because achieving a Tbps data rate with limited antennas is difficult. To address this issue, ELAA technology, comprising hundreds or thousands of antennas, is considered a crucial enabling technology for next-generation communication \cite{ref3}. ELAA enables efficient multiplexing of multiple user equipment (UE) on the same time-frequency resource, thereby improving spectral efficiency and data rates. Additionally, the deployment of ELAA's high beamforming gain facilitates enhanced spatial resolution and compensates for significant path loss experienced in the terahertz frequency bands \cite{ref4}.
\par The high beamforming gain of the ELAA system heavily relies on accurate channel state information (CSI) at the transmitter \cite{ref9}. In the time division duplex (TDD) system, the uplink and downlink exhibit reciprocity, allowing downlink CSI to be obtained through uplink channel estimation. However, in the frequency division duplex (FDD) system, the uplink and downlink operate on different frequencies, weakening the channel reciprocity. As a result, deducing downlink CSI from uplink CSI becomes challenging \cite{ref10}. In other words, CSI can only be acquired through dedicated feedback provided by the UE over signaling channels with limited capacity \cite{ref11}.
\par Currently, there are two typical categories of CSI acquisition methods, the explicit CSI acquisition and the implicit CSI acquisition \cite{ref18}. Explicit feedback schemes directly report an element-wise quantized channel vector. They allow for more flexible transmission or reception methods, which can achieve a higher scheduling gain. Compared to explicit feedback, implicit feedback requires less overhead. Therefore, implicit feedback can enable more accurate link adaptation \cite{ref12}. A mainstream technique in implicit feedback is the codebook-based approach, which feeds back an index of a quantized CSI in a predesign codebook to the transmitter. For codebook-based feedback, the quantization accuracy of CSI depends on the codebook structure and the allowed number of feedback bits \cite{ref13}. The existence of massive antennas in ELAA leads to an unexpected increase in pilot overhead. Therefore, it is crucial to design a codebook to achieve accurate quantization of CSI with limited feedback overhead of the ELAA system.

\subsection{Related Works}
Extensive research has focused on the design of far-field codebooks. The far-field electromagnetic wave can be considered a plane wave, so the phase changes linearly with the antenna index. The 5G new radio (NR) standard adopted a discrete Fourier transform (DFT) codebook for the ULA system, and the two-dimensional DFT (2D-DFT) codebook was introduced for the UPA system \cite{ref14}. Moreover, to enable more accurate CSI acquisition, the NR standard supported codebook oversampling and the linear combination of multiple codewords for feedback \cite{ref15}. IEEE 802.16 m standard adopted an adaptive codebook structure, such as the skewed codebook, and a differential codebook structure, such as a Polar-Cap codebook \cite{ref151}. Besides, in the case of low pilot overhead, the hierarchical codebook \cite{ref15a}, angle-of-departure (AoD) adaptive subspace codebook \cite{ref16}, and compressed sensing (CS) \cite{ref17} methods could also be utilized to quantify CSI accurately. Among them, the codebook feedback schemes based on CS utilize the sparsity of the channel in the angle domain to achieve the goal of low feedback overhead. 

Although the increased antenna numbers of ELAA offer advantages in terms of spectral efficiency and data rates, side effects on wireless channel characteristics brought by it also demand attention \cite{ref5}. The electromagnetic field is generally divided into far-field and near-field, and their boundaries can be determined by the Rayleigh distance $2{D^2}/\lambda$, where $D$ represents the array size and $\lambda$ represents the wavelength \cite{ref6}. Due to the extensive array size and the utilization of high-frequency carriers in the ELAA system, the Rayleigh distance extends to tens or even hundreds of meters. Consequently, User Equipment (UE) is more likely to be positioned within the near-field \cite{ref7}. Unlike the plane wave model of the far-field model, the near-field is usually modeled as a spherical wave \cite{ref8}, \cite{ref9a}. The distance between UE and BS cannot be ignored in the spherical wave model. The far-field codebook has a significant loss when it is used for near-field beamforming. Therefore, the additional distance factor needs to be quantized in the near-field channel, which poses a significant challenge to the design of the near-field codebook.

Only a few studies have focused on codebook design for large near-field ELAA systems. \cite{ref18} designed a codebook for the near-field UPA channel, which uniformly samples in the cartesian coordinates. However, the range of this codebook applicable to near-field channels is limited, so significant quantization errors still exist. To address this issue, \cite{ref20} proposed the sparse polar codebook, sampling in the sparse domain of the polar domain. Furthermore, the codebook was uniformly sampled in the angular domain and non-uniformly in the distance. Besides, \cite{ref19} derived a near-field codebook scheme designing the optimal quantization points based on the Lloyd-Max algorithm. In \cite{ref21}, a hierarchical codebook was designed by projecting the near-field channel into the angle and slope domains, considering the incomplete coverage and overlap of spatial chirp beams, further designing a hierarchical codebook via manifold optimization and alternative minimization.

Unfair sampling methods can render codewords redundant and reduce quantization accuracy. It has been known that the quantization performance of the codebook provides a good indication for the design of codeword sampling. The minimum quantization correlation achieved within the quantization area of the codeword is the pivotal factor determining codeword performance. Remarkably, the quantization performance of the codebook showcases distinct behaviors in the near-field and far-field scenarios. While a sine function characterizes the quantization performance of the codebook for channels in the far-field, research remains absent in evaluating the quantization performance of codewords for near-field channels. Hence, a thorough analysis of near-field codebook quantization performance becomes urgent, given its significance in the design of codebooks. Furthermore, many antennas and the non-negligible distances within the near-field significantly amplify the demand for feedback bits. Consequently, designing a low-quantization bit codebook capable of accurately quantifying the near-field channel is crucial.

\subsection{Contributions}
To fill in this gap, in this paper, we analyze the quantization performance in theory and propose codebook design schemes for the ULA channel and the UPA channel in the ELAA system. Our main contributions are summarized as follows:
\begin{itemize}
    \item [$\bullet$] Firstly, we provide a theoretical analysis of the quantization performance of codeword to channels in the near-field ULA and UPA systems. The quantization performance of the ULA codeword exhibits symmetry and stationary. However, the UPA codeword is non-stationary and asymmetric. Further, we derive a fitting polynomial form of the correlation formula between the codeword and the channel vector. The correlation formula for the ULA codeword can be expressed as an elliptic function, and the fitting correlation formula for each codeword remains constant. In contrast, the correlation formula for the UPA codeword can be represented as an ellipsoid formula, with varying fitting correlation formulas for different codewords. 
    \item[$\bullet$] Secondly, we propose near-field codebook schemes building upon the fitting correlation formula. We present a ULA codebook scheme with uniform sampling in the transform domain that maximizes minimum quantization correlation. Additionally, we introduce an improved dislocation sampling codebook scheme, effectively reducing the overhead of quantized codewords. Recognizing the non-stationarity of UPA channels, we establish the reference ellipsoid as the minimum achievable ellipsoidal shape encompassing the entire quantization area of codewords. Similar to ULA codebook schemes, we develop uniform and dislocation UPA codebook schemes based on the reference ellipsoid. Our analytical results highlight the advantages of oversampling in the angle domain for designing near-field codebooks with high minimum quantization correlation.
    \item[$\bullet$] Lastly, we conduct simulations to compare our proposed codebook with other codebook schemes. The simulation results show that our proposed codebook consistently achieves superior performance compared to other codebook schemes under the same quantization overhead.    
\end{itemize}

\subsection{Organization and Notation}
The remainder of the paper is organized as follows. Section \ref{sec:2} presents the spherical wave models and the CSI quantization feedback model. Section \ref{sec:3} analyzes the characteristic of correlation, and describes the fitting polynomial formula in both ULA and UPA model. In Section \ref{sec:4}, the near-field uniform codebook and dislocation codebook of ULA channel are proposed. Section \ref{sec:5} presents UPA near-field uniform codebook and dislocation codebook. Simulation results are provided in Section \ref{sec:6}, and conclusions are drawn in Section \ref{sec:7}.

\emph{Notations}: Vectors are denoted by lowercase bold letters, while matrices are denoted by uppercase bold letters. $\otimes$ denotes the Kronecker product; $(\cdot)^*$ and $(\cdot)^T$ denotes the conjugate and transpose operations, respectively. 
$\left | \cdot  \right | $ denotes the absolute operator. $\mathrm{diag}(\mathbf{D})$ denotes diagonal matrix from $\mathbf{D}$. $\left \|\mathbf{v }  \right \| $ denotes the Frobenius norm of a vector $\mathbf{v}$. $\left \lfloor x \right \rfloor$ is the rounding symbol.

\section{System Model}
\label{sec:2}
In this section, we first introduce the ULA and UPA spherical wave models of the ELAA system, respectively. Next, we present a CSI quantization feedback model and formulate the design of the codebook as an optimization problem.

\subsection{ULA Near-field Channel Model}

\begin{figure}[t]
    \centering
    \includegraphics[width=     3.5in]{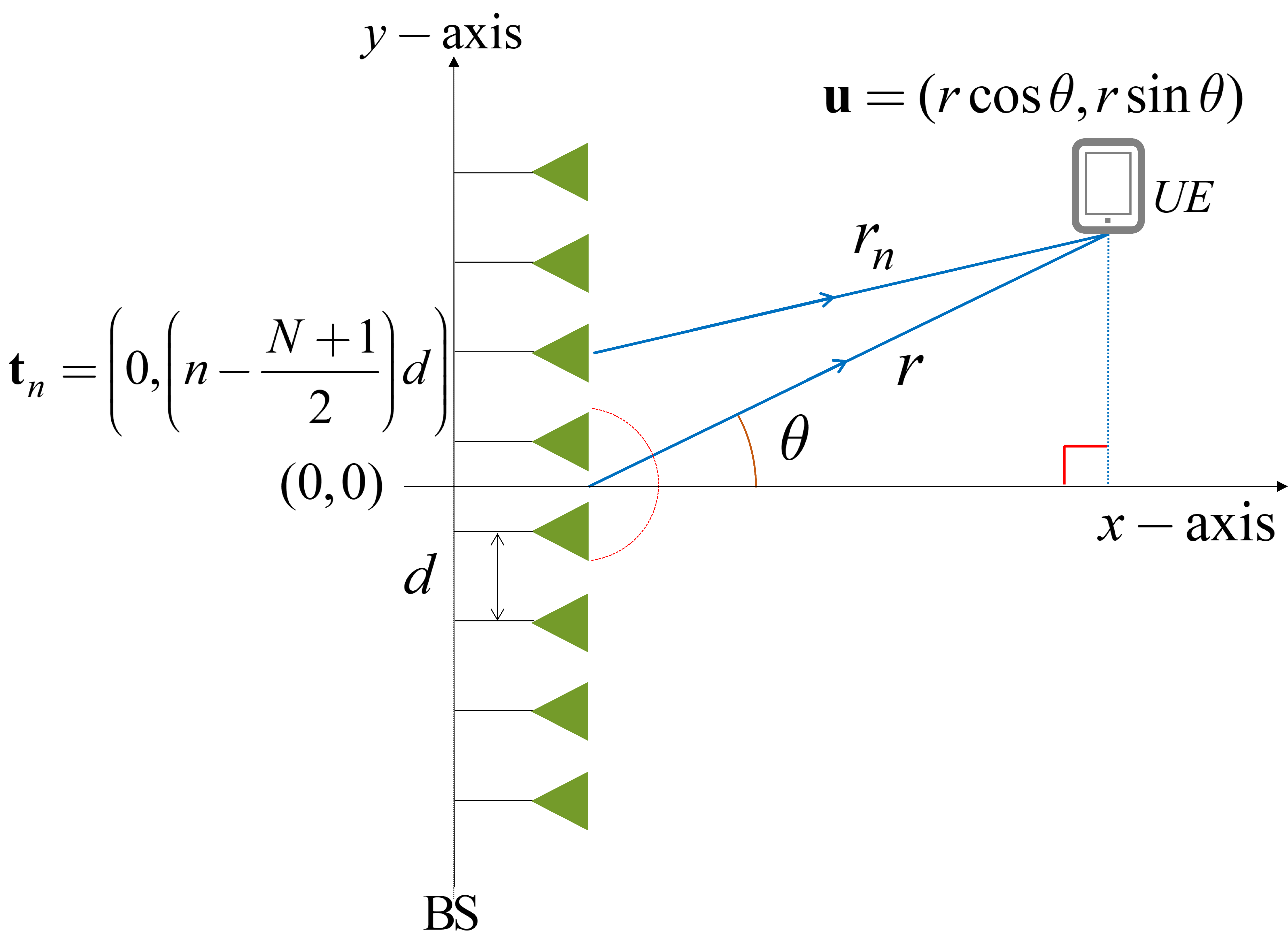}
    \caption{Near-field channel model for ULA communication system.\label{fig1}}
\end{figure}

As shown in Fig. \ref{fig1}, we consider a downlink narrow-band ELAA system, where the BS is equipped with a ULA to serve a single-antenna UE distributed in the near-field region. The $N$-antenna array is placed along the $y$-axis. The antenna spacing is $d = \frac{\lambda}{2}$, where $\lambda$ is the electromagnetic wavelength. The coordinate of the $n$-th antenna is given by $\mathbf{t}_{n} = \left( {0,y_{n}} \right)$, where $y_{n} = \;\left( {n - \frac{N + 1}{2}} \right)d$ with $n = 1,2,\ldots,N$. Meanwhile, the UE is located at $\mathbf{u} = \left( {r{\mathit{\cos}\theta},r{\mathit{\sin}\theta}} \right)$, where $r$ and $\theta$ represent the distance and angle between UE and array center, respectively. 

The line-of-sight (LoS) channel is considered because this paper only focuses on the quantization feedback problem of the near-field codebook. According to the spherical wave model \cite{ref20}, both the angle and distance of UE determine the signal phase, and the near-field channel vector ${\mathbf{h}}$ can be expressed as
\begin{equation}\label{eq1}
{\bf{h}} = \sqrt N g{\bf{b}}\left( {r,\theta } \right),
\end{equation}
where $ k=\frac{2\pi}{\lambda}$ denotes the wavenumber at carrier frequency $f$. $g=\frac{\sqrt \eta e^{ - jkr}}{r}$ represents the complex channel gain, where $\eta$ represents the reference channel gain at a distance of $1 $ m. $\mathbf{b}\left( {r,\theta} \right)$ denotes the near-field beam focusing vector, which is given by
\begin{equation}\label{eq2}
{\mathbf{b}}\left( {r,\theta } \right) = \frac{1}{{\sqrt N }}\left[{e^{ - jk\left( {{r_1} - r} \right)}},{e^{ - jk\left( {{r_2} - r} \right)}}, \ldots ,{e^{ - jk\left( {{r_N} - r} \right)}}\right]^T ,
\end{equation}
where $r_{n} = \left\| {\mathbf{t}_{\mathbf{n}} - \mathbf{u}} \right\|$ represents the distance between the $n$-th antenna at the BS and the UE.
Furthermore, according to the second order Taylor series expansion $\sqrt{1+x} = 1 + \frac{x}{2} - \frac{x^{2}}{8} + \mathcal{O} \left( x^{3} \right)$ , $r_{n}$ can be approximated as
\begin{equation}\label{eq3}
\begin{aligned}
{r_n} &= \sqrt {{{(r\sin \theta - {y_n})}^2} + {{(r\cos \theta )}^2}} \\
& \approx r - \sin \theta {y_n} + \;\frac{{{{\cos }^2}\theta }}{{2r}}y_{n}^{2}.
\end{aligned}
\end{equation}
\begin{remark}\rm
When the $r$ is sufficiently large, the $\frac{\cos^{2}\theta}{2r}$ term can be omitted, and $\mathbf{b}\left( {r,\theta} \right)$ is simplified as
\begin{equation}\label{eq4a}
{\mathbf{a}}\left( \theta\right) = \frac{1}{{\sqrt{N} }}{\left[ {1,{e^{  j\pi \sin \theta }}, \ldots,{e^{ j\pi \left( {N - 1} \right)\sin \theta }}} \right]^T},
\end{equation}
which is equivalent to the conventional far-field beam steering vector for the ULA. In this case, the DFT codebook is adopted to quantify the far-field channel vector. Therefore, to be more precise, the concept of “near-field” in this paper does not exclude far-field as well.
\end{remark}

\subsection{UPA Near-field Channel Model}
\begin{figure}[t]
    \centering
    \includegraphics[width=     3.5in]{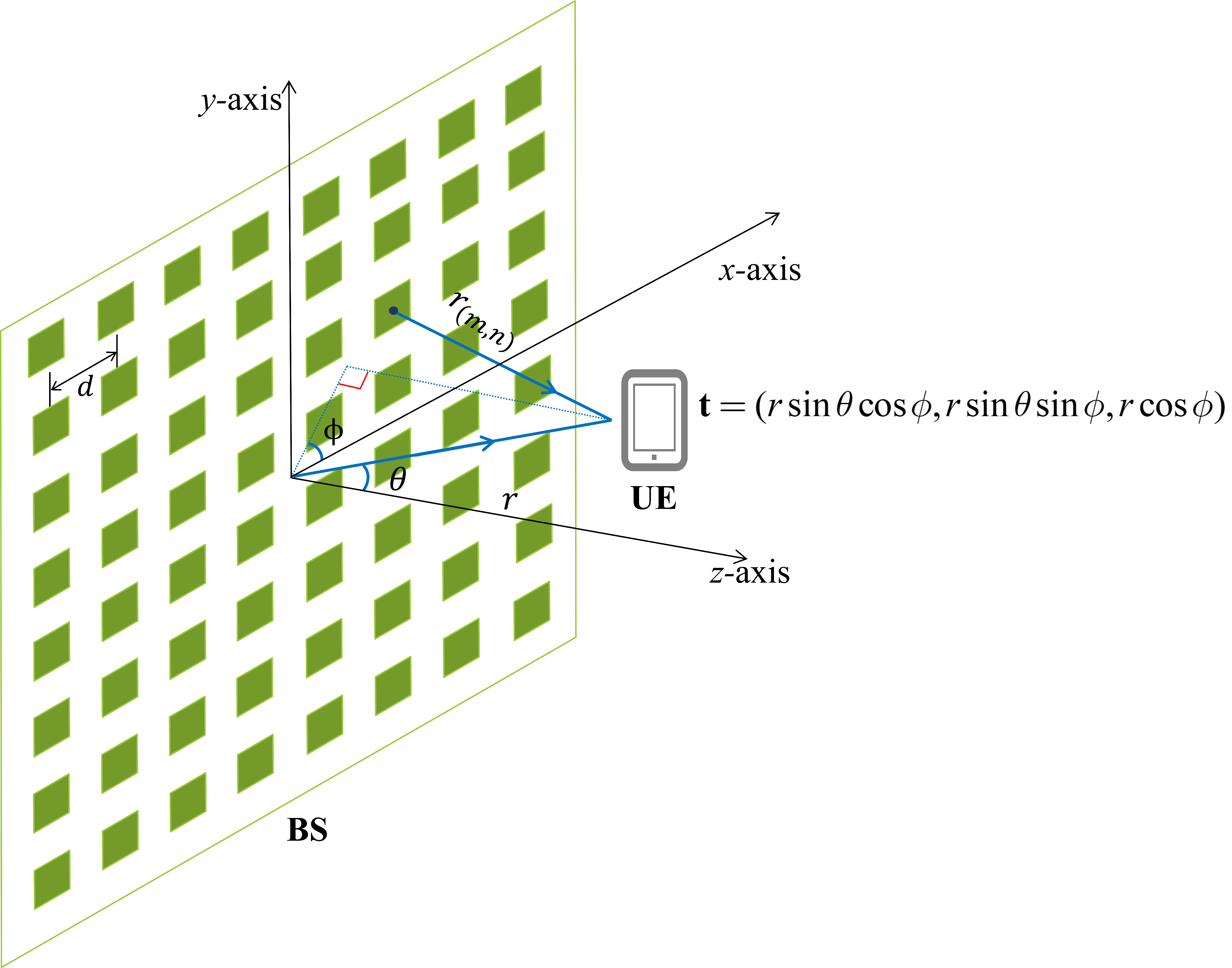}
    \caption{ Near-field channel model for UPA communication system.}
    \label{fig22}
\end{figure}

As shown in Fig. \ref{fig22}, the BS employs a UPA, which is located on the $\rm{xOy}$ plane and the center of the array is located at the coordinate origin. $N \times N$ uniformly spaced antenna elements are placed in both horizontal and vertical directions, with a spacing of $d = \frac{\lambda}{2}$. The Cartesian coordinate of the $(m,n)$-th antenna element of the UPA can be expressed as $\mathbf{t}_{(m,n)} = \left( {x_{m},y_{n},0} \right)$ with $x_{m} = \left( {m - \frac{N + 1}{2}} \right)d$, $y_{n} = \left( {n - \frac{N + 1}{2}} \right)d$, $m = 1,...,N$, $n = 1,\ldots,N$. Meanwhile, we assume the coordination of UE is ${\bf{u}} = \left( {r\sin \theta \cos \phi,r\sin \theta \sin \phi ,r\cos \theta } \right)$, where $r$, $\theta$ and $\phi$ represent the distance, elevation angle and azimuth angle of UE relative to the UPA center, respectively. Therefore, the beam focusing vector for UPA can be obtained based on the spherical wave propagation model as
\begin{equation}\label{eq5}
{\mathbf{b}}\left( {r,\theta ,\phi } \right) = \frac{1}{{ N }}{\left[{e^{ - jk\left( {{r_{\left( {1,1} \right)}} - r} \right)}},  \ldots ,{e^{ - jk\left( {{r_{\left( {N,N} \right)}} - r} \right)}}\right]^T},
\end{equation}
where $r_{({m,n})} = \left\| {\mathbf{t}_{({m,n})} - \mathbf{u}} \right\|$ represents the distance between the $(m,n)$-th antenna at the BS and the UE, which can be approximated as
\begin{eqnarray}\label{eq6}
\begin{aligned}
r_{(m,n)} &\approx  r - \sin\theta \cos\phi {x_m} - \sin\theta \sin\phi {y_n} \\
& \quad + \frac{{1 - {{\sin }^2}\theta {{\cos }^2}\phi }}{{2r}}x_m^2 + \frac{{1 - {{\sin }^2}\theta {{\sin }^2}\phi }}{{2r}}y_n^2\\
& \quad- \frac{{{{\sin }^2}\theta \cos\phi \sin\phi }}{r}{x_m}{y_n}.
\end{aligned}
\end{eqnarray}

\begin{remark}
\rm
When the $r$ is sufficiently large, the last 3 terms in (\ref{eq6}) can be omitted, and $\mathbf{b}\left( {r,\theta,\phi} \right)$ is simplified as 
\begin{equation}\label{eq7a}
\begin{split}
\mathbf{a} \left( \theta, \phi \right)  = & \frac{1}{ N } \left [ 1,\ldots,e^{  j\pi (m \sin\theta \cos\phi+ n\sin\theta \sin\phi) }, \dots, \right.  \\
& \left. \quad e^{ j\pi((N-1)\sin\theta \cos\phi + (N-1)\sin\theta \sin\phi )} \right ]^T,
\end{split}
\end{equation}
which is equivalent to the conventional far-field beam steering vector for the UPA, and the 2D-DFT codebook is adopted for the CSI feedback. Since the phase of (\ref{eq7a}) can be decoupled into two parts in terms of $x$ and $y$, the 2D-DFT codebook can be expressed in the form of the Kronecker product of the DFT vectors, that is 
\begin{equation}
{\bf{a}} = {{\bf{a}}_x} \otimes {{\bf{a}}_y},
\end{equation}
where ${{\bf{a}}_x} = \frac{1}{N}\left[ {1,{e^{  j\pi \sin \theta \cos \phi }}, \ldots ,{e^{  j\pi (N - 1)\sin \theta \cos \phi }}} \right]$ and ${{\bf{a}}_y} = \frac{1}{N} \left[ {1,{e^{ j\pi \sin \theta \sin \phi }}, \ldots ,{e^{ j\pi (N - 1)\sin \theta \sin \phi }}} \right]$. However, the cross-term $\frac{{{{{\sin }^2}\theta \cos\phi \sin\phi }}}{r}{x_m}{y_n}$ in (\ref{eq6}) prevents it from being decoupled as $\mathbf{a}\left( {\theta,\phi} \right)$. 
\end{remark}

\subsection{CSI Quantization Feedback Model}
For FDD communication systems, the BS can obtain the CSI through the UE’s feedback. Specifically, the pilot signal $\mathbf{X} = \text{diag}\big( {x_{1},\ldots,x_{\widetilde{N} }} \big)$ is transmitted at first, where $x_{n}$ denotes the $n$-th pilot symbol, and $\widetilde{N} $ represents the size of either ULA or UPA. At the UE side, the received signal is given by
\begin{equation}\label{eq8}
{\mathbf{y}} = {\mathbf{X}\bf{h}} + {\mathbf{n}},
\end{equation}
where $\mathbf{n}$ is the additive Gaussian white noise (AWGN) with variance $\sigma^{2}$. Based on this, the estimation of channel vector, denoted as $\tilde{\mathbf{h}}$, can be obtained by methods such as least squares (LS) \cite{ref22}. Since this paper mainly focuses on the codebook design for the near-field communication, the perfect CSI estimation is assumed, i.e., $ \tilde{\mathbf{h}} = \mathbf{h}$. 

To inform the BS of CSI with limited feedback, the channel vector is quantized based on a predefined codebook $\mathbf{W} = \left[\mathbf{w}_{1},\ldots,\mathbf{w}_{S} \right]$, which contains $\mathrm{S}$ codewords and satisfies $\left\| \mathbf{w}_{s}\right\| = 1$. The UE selects the optimal codeword from the codebook and feeds back its index $s^\mathrm{\star } =\arg\underset{s}{\max}\left| {\mathbf{h}^{T}\mathbf{w}_{s}} \right|^{2}$ to the BS. Subsequently, the BS can determine the transmission scheme based on the CSI feedback from the UE. For instance, the codeword $\mathbf{w}_{s^\mathrm{\star}}$ can be utilized as the beamforming weight.

In the above quantization feedback model, the codebook design influences the accuracy of channel quantization, which in turn affects the performance of the communication system. Obviously, all the channel vectors in the area of interest have different correlations with the codewords. This paper adopts the max-min correlation criterion, and assuming $\mathcal W =\left \{ \mathbf{w}_{1}, \dots,\mathbf{w}_{S} \right \}$, the problem of near-field codebook design can be formulated as follows
\begin{eqnarray}\label{eq9}
\begin{aligned}
\mathop {{\rm{max}}}\limits_{\mathbf{W}} & {\mkern 1mu} \mathop {{\rm{min}}}\limits_{{\bf{h}} \in {\cal H}} {\mkern 1mu} \mathop {{\rm{max}}}\limits_s |{{\bf{h}}^{\rm{T}}}{{\bf{w}}_s}|\;\\
{\rm{s}}{\rm{.t}}{\rm{. }}& \left| {\mathcal{W}} \right| = S.
\end{aligned}
\end{eqnarray}
or
\begin{eqnarray}\label{eq9a}
\begin{array}{l}
\mathop {\min }\limits_{\mathcal{W}} \;{\mkern 1mu} \left| {\mathcal{W}} \right|\\
{\rm{s}}{\rm{.t}}{\rm{.}}\mathop {\;{\rm{min}}}\limits_{{\bf{h}} \in {\cal H}} {\mkern 1mu} \mathop {{\rm{max}}}\limits_s {\mkern 1mu} |{{\bf{h}}^{\rm{T}}}{{\bf{w}}_s}| > c. 
\end{array}
\end{eqnarray}
${\cal H}$ represents the set of LoS channel vectors within the area of interest, and $c\in(0,1)$ represents the correlation between codewords and channel vectors. 

For the codebook design issue, existing solutions include
Grassmannian codebooks \cite{ref23}, random vector quantization (RVQ) codebooks \cite{ref24}, and generalized Lloyd codebooks \cite{ref25}. Nevertheless, these methods cannot fully utilize the characteristics of near-field channels while guaranteeing compatibility with the existing far-field codebooks. Therefore, we consider select codewords from beam focusing vectors $\mathbf{b}(r,\theta)$, and the existing DFT codebook for the far-field is based on this idea.

\section{Codeword Quantization Performance Analysis}
\label{sec:3}
This section first investigates the correlation between the near-field codewords and the channel vectors. The transform domain perspective for analyzing correlation function is proposed, demonstrating many desired mathematical properties. Secondly, the section provides a fitting formula for the correlation quantization performance of the ULA and UPA codewords, which serves as inspiration for codebook design.
\subsection{Correlation Function for ULA systems}
Codewords can be selected from the beam focusing vectors, i.e., ${\mathbf{w}_{s}} = {\mathbf{b}}^*( {r_s,\theta_s } )$, they can be viewed as LoS channel vectors at specific positions. 
In the ULA systems, the correlation between the codeword $\mathbf{w}_s$ and the normalized channel vector pointing to $({{r_q},{\theta _q}})$ can be calculated as 
\begin{equation}\label{eq10}
    \begin{aligned}
        \tau \big( {{r_s},{\theta _s}; {r_q},{\theta _q}}\big)
        = &\big|{{{\mathbf{b}}}\big({{r_q},{\theta _q}}){\mathbf{b}^*}( {{r_s},{\theta _s}}\big)}\big|\\
        = & \frac{1}{N}\Bigg| \sum\limits_{n = 1}^N {\exp }\Bigg( - j\frac{2\pi }{\lambda }\bigg(\big( {\sin {\theta _q} - \sin {\theta _s}}\big){y_n} \\
        &+\Big( {\frac{{{{\cos }^2}{\theta _s}}}{{2{r_s}}} - \frac{{{{\cos }^2}{\theta _q}}}{{2{r_q}}}}\Big) y_n^2\bigg) \Bigg)\Bigg|.\\
    \end{aligned}
\end{equation}
Let $\alpha_i=\frac{\lambda \cos^2\theta_i}{4r_i}$ and $\beta_i =\sin\theta_i$ with $i=s,q$. $\delta _{\alpha } $ and $\delta _{\beta } $ respectively represent as the position difference between $\mathbf{h}=\mathbf{b}(r_{q}, \theta_{q})$ and $\mathbf{w}_s$, which can be expressed as
\begin{equation}
    \begin{split}
      &\delta_{\alpha } =\alpha_{q}-\alpha_{s},\quad \delta _{\beta }=\beta_{q}-\beta_{s}.
    \end{split}
    \label{replace}
\end{equation}
Then, (\ref{eq10}) can be simplified as
\begin{equation} \label{eq12}
\begin{split}
f&\left(\delta_{\alpha },\delta_{\beta}\right)\\
&=\!\frac{1}{N} \Bigg|\sum_{n=1}^{N}\exp\Bigg(\!-j\pi \bigg(\!-\delta_{\alpha}n^2  +\Big(\delta_{\beta}+\delta_{\alpha} \big(N+1\big)\Big)n\bigg)\Bigg)\Bigg|.
\end{split}
\end{equation} 
Without loss of generality, the correlation between the codeword and the channel vector always satisfies {$f\left(\delta_{\alpha } ,\delta_{\beta } \right)\le 1$}. The condition {$f\left(\delta_{\alpha}, \delta_{\beta}\right) = 1$} holds if and only if {$\delta_{\alpha} = 0$} and {$\delta_{\beta} = 0$}. Consequently, a quantization error is always present when a codeword quantizes a channel other than itself.

\begin{figure}
  \centering
\includegraphics[width=3.5in]{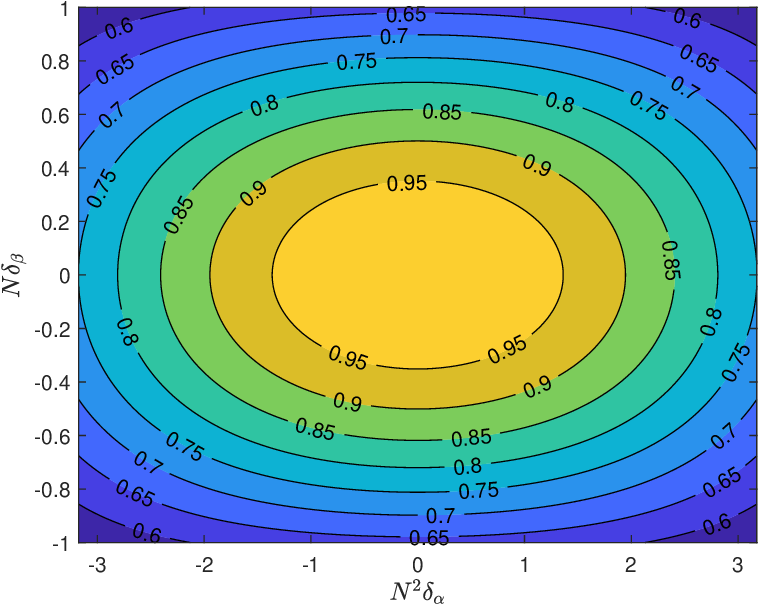}
  \caption{Contour distribution between codeword quantization correlation and position difference with $f=100\mathrm{GHz}$ and $N=512$.}
  \label{fig2}
\end{figure}

To further explore the characteristics of quantization performance, we perform a normalization substitution for the variables in the above equation. We set ${\tilde \delta _{\alpha }  }\! = \!{\delta_{\alpha}}{N^2}$, ${\tilde \delta _{\beta }  }\! = \!{\delta_{\beta}}{N}$ 
and $t \! =\! \frac{n}{N-1}\!-\!\frac{1}{2}$. 
For the antenna with sufficiently large of $N$, the above formula can be approximated as
\begin{equation}\label{eq13}
\tilde{f}\left({\tilde\delta_{\alpha }},{ {\tilde \delta _{\beta}  }} \right) \approx \frac{1}{N} \left| {\int_{ - 1/2}^{1/2} {\exp } \left( { - j\pi \left( {{{\tilde \delta _{\beta }  }}t - {{\tilde \delta _{\alpha }  } }{t^2}} \right)} \right)dt} \right|.
\end{equation} 
We plot the graph of $\tilde{f}\left({\tilde\delta_{\alpha }},{ {\tilde \delta _{\beta}  }} \right)=c$ in Fig. \ref{fig2}, where $c\in(0,1)$. Interestingly, the boundaries of the codeword quantization areas resemble ellipses. The quantization performance remains independent of the frequency. The codeword quantization regions are solely determined by the number of antennas and the minimum correlation $c$. When $c$ is constant, the coverage area of codewords is inversely proportional to $N$ for the $\alpha$ domain and $N^2$ for the $\beta$ domain.


Unfortunately, the exponential term in (\ref{eq12}) is relatively complicated. Most existing literature deals with the exponential term based on the Fresnel integral, which is still difficult to directly analyze the quantification performance of codewords \cite{ref20}. To solve the problem, we will give an approximate fitting correlation formula.
Before that, we first explore the correlation properties of codewords in the following.
\begin{property}[Stationary]\label{P11}
 \rm
The correlation between the codeword and channel vector is only related to {$\delta _{\alpha } $} and {$\delta _{\beta } $}, but it is independent of the codeword. For different codeword {$\mathbf{w}_s$} and {$\mathbf{w}_{s^{'}}$}, the quantization performance for the channel vector within its quantization area is always the same. The stationarity in the ULA channel can be formulated as 
\begin{equation}\label{eq14}
\begin{split}
f(\alpha_s,\beta_s;\alpha_s\!+\!\delta _{\alpha } ,\beta_s\!+\!\delta _{\beta })\!=\!f\left(\alpha_{s^{'}},\beta_{s^{'}};\alpha_{s^{'}}\!+\!\delta _{\alpha } ,\beta_{s^{'}}\!+\!\delta_{\beta }\right).
\end{split}
\end{equation}
\end{property}
\begin{proof}
\rm
See Appendix \ref{proof1}.
\QEDA
\end{proof}
\begin{property}[Symmetric]\label{pro1}
\rm
The correlation distribution of the near-field channel is symmetric. Within the quantization area of the codeword, the correlation between the codeword and the channel vectors is symmetric about the codeword in the $\alpha $ and $\beta$ domain, which can be expressed as
\begin{equation}\label{eq15}
\begin{split}
f\left(\delta _{\alpha } ,\delta _{\beta } \right)=f\left(\delta _{\alpha } ,-\delta _{\beta } \right)&=f\left(-\delta _{\alpha } ,\delta _{\beta } \right)=f\left(-\delta _{\alpha } ,-\delta _{\beta } \right).
\end{split}
\end{equation}
\end{property}
\begin{proof}
\rm
See appendix \ref{proof2}.
\QEDA
\end{proof}

Inspired by (\ref{eq12}) and the above properties, we use a polynomial function to approximate the correlation function in Corollary \ref{Propo1}.  

\begin{corollary}\label{Propo1}
\rm
For any ULA codeword $\mathbf{w}_s=\mathbf{b}^*(\alpha_s,\beta_s)$, the fitting polynomial formula of quantization performance {$f(\delta _{\alpha } ,\delta _{\beta } )$} can always be expressed as 
\begin{equation}
f\left( {{\delta _{\alpha }  },{\delta _{\beta }  }} \right) \approx {p_\alpha }{\delta _{\alpha } }^2{N^4} + {p_\beta }{\delta _{\beta }  }^2{N^2} + 1 ,
\end{equation}
where 
\begin{equation}\label{eq18}
\begin{split}
p_\alpha=-0.025983670363830,\\
p_\beta=-0.391749735984250. 
\end{split}
\end{equation}
To conclude, for the ULA codeword {$\mathbf{w}_s$} with a minimum quantization correlation of $c$, the distribution of $\delta _{\alpha }$ and $\delta _{\beta }$ that satisfies $f(\delta _{\alpha },\delta _{\beta }) =c$ can be considered equivalent to
\begin{equation}\label{eq19}
    p_\alpha\delta _{\alpha }^{2} N^4+ p_\beta\delta _{\beta }^{2}N^2=c-1.
\end{equation}
This formula can be further simplified into the form of the following formula
\begin{equation}\label{eq20}
    \frac{p_\alpha\delta _{\alpha }^{2}N^4}{c-1}+ \frac{p_\beta \delta _{\beta }^{2}N^2}{c-1} =1.
\end{equation}
\end{corollary}

Evidently, the correlation fitting formula in Corollary \ref{Propo1} is an elliptic function. The ellipse formula always centers around $(\alpha_s,\beta_s)$. And the ellipse is the quantization boundary of {$\mathbf{w}_s$} when the minimal quantization correlation is {$c$}. The minimal quantization correlation affects the major and minor axes of the ellipse formula. The axial length of the ellipsoid and the area of codeword quantization decrease with the increase of $c$. 
Further, the axis length of the ellipse is also decided by {$N$}. The axis length of the ellipse in the {$\alpha $} and {$\beta $} domains are inversely proportional to {$N^2$} and {$N$}, respectively. 

The quantization performance of ULA codeword $\hat{\mathbf{w}}=\mathbf{b}^*(0,0)$ can be written as {$f(\alpha_q,\beta_q)$}. The possible channel vectors always distribute in the ellipse interior with the minimum correlation as $c$, which can be formulated as
\begin{equation}\label{eq21}
    \Omega =\left \{ \mathbf{b}\left(\alpha_q,\beta _q\right)\Big|\frac{{p_\alpha N^4}\alpha_{q}^{2}}{{c-1}}+\frac{{p_\beta N^2}\beta_{q}^{2}}{{c-1} } \leq 1\right \}.
\end{equation}
With stationarity in the ULA system, the quantization area of any codeword can be represented using ellipses with the same axis length as $\hat{\mathbf{w}}$ but different centers.
The formula in Corollary \ref{Propo1} is concise and provides strong theoretical support for the codebook design scheme outlined in this paper. 

\begin{figure}
  \centering
 \includegraphics[width=3.5in]{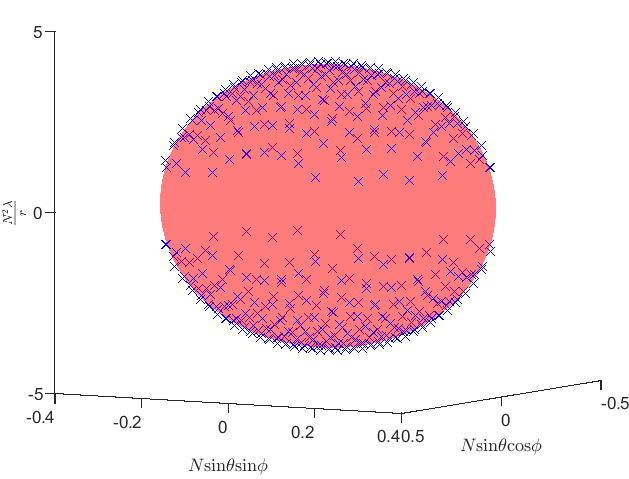}
  \caption{The fitting quantization area and actual channel vector of $\check{\mathbf{w}}$ with $c=0.95$, $f=100$ GHz and number of antenna $16\times16$.}
  \label{fig3a}
\end{figure}

\subsection{Correlation Function for UPA systems}
In the UPA system, the correlation between codeword $\mathbf{w}_s=\mathbf{b}^*(r_{s},\theta_{s},\phi_{s})$ and channel vector pointing to $(r_{q},\theta_{q},\phi_{q})$ can be calculated as 
\begin{equation}\label{eq22}
 \tau \left({r_q},{\theta _q},{\phi _q};{r_s},{\theta _s},{\phi _s}\right)  = \left| {{{\bf{b}}}\left( {{r_s},{\theta _s},{\phi _s}} \right){\bf{b}^*}\left( {{r_q},{\theta _q},{\phi _q}} \right)} \right|.
\end{equation}
Let $\psi_i=\sin\theta_i\cos\phi_i$, $\varphi_i=\sin\theta_i\sin\phi_i$ and $\rho_i =\frac{\lambda}{r_i}$ with $i=s,q$. We replace the difference of position between the codeword and channel vector as follows 
\begin{equation}\label{eq23}
\begin{split}
\delta_{\psi_s}&=\sin\theta_q\cos\phi_q-\sin\theta_s\cos\phi_s, \\
\delta_{\varphi_s}&=\sin\theta_q\sin\phi_q-\sin\theta_s\sin\phi_s, \\
\delta_{\rho_s} &=\frac{\lambda}{r_q}-\frac{\lambda}{r_s}.
\end{split}
\end{equation}
(\ref{eq22}) can be rewritten as (\ref{eq23a}).
\newcounter{TempEqCnt} 
\setcounter{TempEqCnt}{\value{equation}} 
\begin{figure*}[t]
\begin{equation}
\label{eq23a}
\begin{split}
f&(\psi_s, \varphi_s, \rho_s; \delta_{\psi_s}, \delta_{\varphi_s}, \delta_{\rho_s}) \\
=&\frac{1}{N^2} \bigg| \sum_{n=1}^{N} \sum_{m=1}^{N} \exp \bigg(-j\Upsilon _{(m,n)}\bigg(\psi_s, \varphi_s, \rho_s;\delta_{\psi_s},\delta_{\varphi_s},-\delta_{\rho_s}\bigg)\bigg)\bigg| \\
=&\frac{1}{N^2} \bigg| \sum_{n=1}^{N} \sum_{m=1}^{N} \exp \bigg(-j\pi \bigg(m \delta_{\psi_s}\! +\! n \delta_{\varphi_s}\!-\!\frac{\rho_s(1\!-\! \psi_s)^2\!-\!(\rho_s\!-\!\delta_{\rho_s})(1\!-\!\psi_s\!+\!\delta_{\psi_s})}{4} \bigg(m\!-\!\frac{N+1}{2}\bigg) ^2-\bigg(\frac{\rho_s(1\!-\!\varphi_s)^2\!}{4} \\
&-\frac{ \!(\rho_s-\delta_{\rho_s})(1\!-\!\varphi_s\!+\!\delta_{\varphi_s}\!)}{4}
\bigg)\bigg(n\!-\!\frac{N\!+\!1}{2}\bigg)^2\! +\! \frac{\rho_s\varphi_s\psi_s\!-\! (\psi_s\! - \! \delta_{\psi_s})(\varphi_s \! -\! \delta_{\varphi_s})(\rho_s \! -\! \delta_{\rho_s})}{2}\bigg(n\!-\! \frac{N+1}{2}\bigg)\bigg(m\!-\!\frac{\!N+\!1}{2}\bigg)\bigg)\bigg)\bigg|.
\end{split}
\end{equation}
\hrulefill
\vspace*{10pt}
\end{figure*}

The cross-term {$x_my_n$} contained in the UPA channel vector prevents the use of the Kronecker product to decouple the channel vector. Ignoring the cross-term {$x_my_n$} would result in significant performance errors due to the loss of crucial CSI. Before giving the method to solve the problem, the characteristics of UPA channel correlation shown in the Property \ref{pro3} and Property \ref{pro4} are discussed. 
\begin{property}[Non-stationary]\label{pro3}
\rm
The quantized areas of different UPA codewords are not consistent under the same minimum correlation {$c$}. For two UPA codewords $\mathbf{w}_s=\mathbf{b}^*(\psi_s,\varphi_s,\rho_s)$ and $\mathbf{w}_{s^{'}}\!=\!\mathbf{b}^*(\psi_{s^{'}},\varphi_{s^{'}},\rho_{s^{'}})$, the non-stationary feature can be expressed as
\begin{equation}\label{eq25}
\begin{split}
f\!\Big(\!\psi_s,\varphi_s,\rho_s;\delta_{\psi},\delta_{\varphi},\delta_{\rho}\Big)\!\ne\!f\!\left(\psi_{s^{'}},\varphi_{s^{'}},\rho_{s^{'}};\delta_{\psi},\delta_{\varphi},\delta_{\rho}\!\right)\!.
\end{split}
\end{equation}
\end{property}

\begin{proof}
See Appendix \ref{proof3}.
\QEDA
\end{proof}

\begin{property}[Asymmetrical]\label{pro4}
\rm
For the UPA codeword $\mathbf{w}_s=\mathbf{b}^*(\psi_s,\varphi_s,\rho_s)$ , the quantization performance of the codeword is asymmetrical, which can be expressed as
\begin{equation}\label{eq26ab}
\begin{split}
f\Big(\psi_s,&\varphi_s,\rho_s;\psi_s\!+\!\delta_{\psi_s},\varphi_s\!+\!\delta_{\varphi_s},\rho_s\!+\!\delta_{\rho_s}\Big)\\&\ne\!f\left(\psi_{s},\varphi_{s},\rho_{s};\psi_{s}\!+\kappa_{\psi} \delta_{\psi_s},\varphi_{s}\!+\kappa_{\varphi} \delta_{\varphi_s},\rho_{s}\!+\kappa_{\rho} \delta_{\rho_s}\right), 
\end{split}
\end{equation}
where $\kappa_{\psi}$, $\kappa_{\varphi}$, and $\kappa_{\rho}$ are equal to 1 or -1 but will not be equal to 1 at the same time. 
\end{property}

\begin{proof}
See Appendix \ref{proof4}.
\QEDA
\end{proof}

Unlike the ULA model, the non-stationarity of the UPA model poses a challenge for designing UPA codewords. It should be noted that $\iota ^{(\rho)}_{(m,n)} \approx 0$ at high frequencies. Therefore, in the following, we assume that UPA codewords exhibit symmetry at high frequencies. Simulation results demonstrate the reasonableness of our assumptions. In Corollary \ref{Propo2}, we provide the fitting polynomial formula for the UPA codewords quantization performance.

\begin{corollary}\label{Propo2}
\rm
For a UPA codeword $\mathbf{w}_s=\mathbf{b}^*(\psi_s,\varphi_s,\rho_s)$,
a polynomial function can be used to better fit the correlation between $\mathbf{w}_s$ and the UPA channel vector $\mathbf{w}_q=\mathbf{b}^*(\psi_q,\varphi_q,\rho_q)$, and the fitting formula can be written as
\begin{equation}\label{eq27}
\begin{split}
f\Big(\psi_s,&\varphi_s,\rho_s;\psi_s\!+\!\delta_{\psi_s},\varphi_s\!+\!\delta_{\varphi_s},\rho_s\!+\!\delta_{\rho_s}\Big)\\
& =p_{\psi_s}\delta _{\psi _{s} }^{2} N^2+p_{\varphi_s}\delta _{\varphi _{s} }^{2} N^2+p_{\rho_s}\delta _{\rho _{s} }^{2}N^4+1.
\end{split}
\end{equation}
Due to the non-stationarity of the UPA channel, it should be noted that the parameters $p_{\psi_s}$, $p_{\varphi_s}$, and $p_{\rho_s}$ of the fitting formula are closely related to the quantization center of the codeword $\mathbf{w}_s$.  With minimum correlation of $c$, the above formula can be transformed into the following ellipsoid formula
\begin{equation}\label{eq28}
  \frac{{p_{\psi_s} N^2}\delta _{\psi _{s} }^{2} }{{c-1} }+ \frac{{p_{\varphi_s} N^2}\delta _{\varphi _{s} }^{2} }{{c-1} }+ \frac{{p_{\rho_s} N^4}\delta _{\rho _{s} }^{2} }{{c-1} }=1.
\end{equation}
\end{corollary}

The correlation fitting formula in Corollary \ref{Propo2} provides the quantization boundary of the codeword {$\mathbf{w}_s$} with the minimum correlation {$c$}. The fitting formula has different axial lengths for different codewords due to the non-stationary feature. The fitting formula in Corollary \ref{Propo2} offers strong support for the codebook design of near-field UPA channel.
 
With a minimum correlation of $c$, the quantized channel vectors of codeword {$\check{\mathbf{w}}\!=\!\mathbf{b}^{*}(0,0,0)$} are always distributed within the ellipsoid interior. This can be formulated as
\begin{equation}\label{eq29}
\begin{split}
\Omega =\left \{\mathbf{b}\left(\psi_q,\varphi_q,\rho_q\right)\Big|
\frac{p_{\psi_s} N^2\psi_{q}^{2} }{c-1} \!+ \!
\frac{p_{\varphi_s} N^2\varphi _{q}^{2} }{c-1}  \!+  \!
\frac{p_{\rho_s} N^4 \rho _{q}^{2} }{c-1}  \! \leq  \! 1\right\}.
\end{split}
\end{equation}
The above ellipsoid with $c=0.95$ is depicted in Fig. \ref{fig3a}.
We also plot the actual UPA channel vector $\mathbf{w}(\psi_q,\varphi_q,\rho_q)$, which satisfies the condition {$f(0,0,0;\psi_q,\varphi_q,\rho_q)=0.95$}. It can be observed that the actual channel vectors are consistently distributed on the fitting ellipsoid. This example verifies the accuracy of the fitting ellipsoid formula in Corollary \ref{Propo2}.

\section{ Near-field ULA Codebook Design}
\label{sec:4}
We have analyzed quantization performance and proposed the correlation fitting formula of codewords. This section proposes two ULA near-field codebook schemes, that achieve the maximum quantization area under a fixed minimum correlation. First, the uniform codebook with uniform intervals is proposed. In addition, we redesign a codebook scheme with dislocation sampling points to further reduce the quantization overhead.
 
\subsection{Uniform Quantization Codebook Scheme }
\begin{figure}
  \centering
  \subfloat[]{
  \label{fig88}
  \includegraphics[width=1.72in]{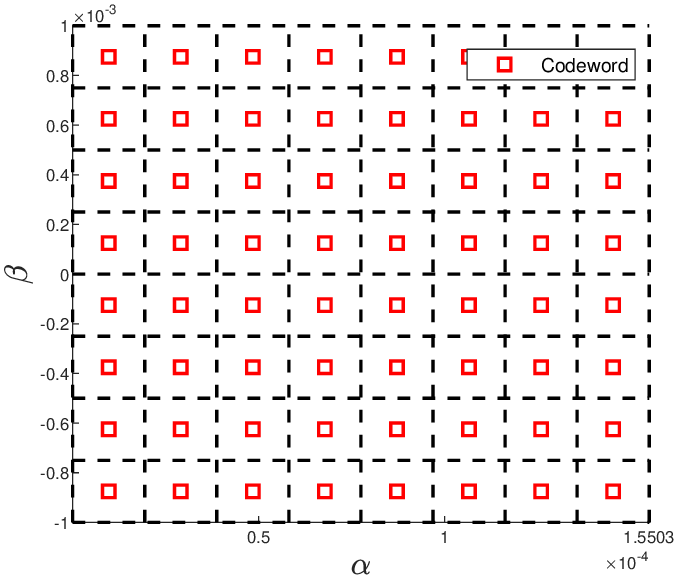}
  }
  \subfloat[]{
  \label{fig7}
  \includegraphics[width=1.72in]{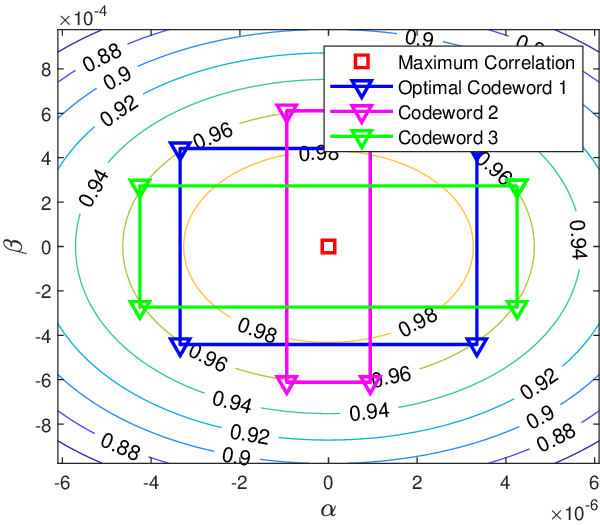}}

  \caption{ULA uniform codebook: (a) Resource division in the $\alpha$-$\beta$ domain. (b) Quantization area of a codeword with minimum correlation $c$.}
  \label{fig:ENISA1}
\end{figure}

The most commonly adopted approach to obtain the quantified positions of the codewords is to perform uniform sampling in the $\alpha$-$\beta$ domain, as shown in Fig. \ref{fig:ENISA1}\subref{fig88}. And Fig. \ref{fig:ENISA1}\subref{fig7} presents the quantification performance of the codeword $\hat{\mathbf{w}}={\mathbf{b}}(0,0)$. The quantization area of the codeword is rectangular. A red square at the center of the rectangular quantization area represents the codeword. Moreover, each blue triangle represents the intersection of the quantization intervals of adjacent codewords. Multiple layouts of rectangular vertices on the quantization boundary constitute multiple quantization schemes. In order to improve the quantization accuracy of the codeword, it is always desired that the codeword has the largest quantization area under minimum quantization correlation $c$. Therefore, the goal can be mapped as maximizing the area of the rectangle on a specific ellipse.

The quantization performance of the codeword $\hat{\mathbf{w}}$ can always be represented by the (\ref{eq20}). Consider a vertex $(\alpha^\star,\beta^\star)$ of the rectangle in Fig. \ref{fig:ENISA1}\subref{fig7}, where $\beta^\star>0$ and $\alpha^\star>0$. 
According to the Cauchy-Schwartz inequality, for any point on the inscribed rectangle of an ellipse, the area of the rectangle is the largest when the point locates at $\bigg(\frac{1}{N^{2}}\sqrt{\frac{\left( {c - 1} \right)}{2p_{\alpha}}}, \frac{1}{N}\sqrt{\frac{\left( {c - 1} \right)}{2p_{\beta}}}\bigg)$.
Therefore, when the channel correlation is $c$, the sampling interval for achieving the maximum quantization area on the $\alpha$-$\beta$ domain can be calculated as
\begin{equation}\label{fig5}
\Delta \alpha \! = \! 2\alpha^\star \!=\!\frac{1}{N^{2}}\sqrt{\frac{2\left( {c - 1} \right)}{p_{\alpha}}}, \quad \Delta \beta \!= \! 2\beta^\star\!=\! \frac{1}{N}\sqrt{\frac{2\left( {c - 1} \right)}{p_{\beta}}}.
\end{equation}
Consider that the UE distribution is within a range of distances $r \in \left[0.62\sqrt{\frac{D^{3}}{\lambda}},\infty\right)$ and angles $\theta \in \left[-\frac{\pi }{2},\frac{\pi }{2}\right]$. The maximum quantization range on the $\alpha$ and $\beta$ domain can be calculated as
\begin{equation}\label{eq31}
    Q_{\alpha} = \sqrt{\frac{\lambda}{2.48D^{3}}} \approx \frac{1}{N\sqrt{N}},\quad Q_{\beta} = 2.
\end{equation}
Then, the number of codewords in the $\alpha$ domain and $\beta$ domain are given by
\begin{equation}\label{eq32}
    S_{\alpha} = \frac{Q_{\alpha}}{\Delta{\alpha}} = \sqrt{\frac{Np_{\alpha}}{2\left( {c - 1} \right)}} , \quad S_{\beta} = \frac{Q_{\beta}}{\Delta{\beta}} =N\sqrt{\frac{2p_{\beta}}{\left( {c - 1} \right)}} .
\end{equation}
Thus, the total number of codewords to achieve the minimum number of feedback bits can be calculated as
\begin{equation}\label{eq33}
   S_{ULA} =S_{\alpha } S_{\beta}  = \frac{N\sqrt{Np_{\alpha}p_{\beta}}}{\left( {1 - c} \right)}.
\end{equation}
The $s_{\alpha}$-th sampling points in the $\alpha$ domain can be reformulated as
\begin{equation}\label{eq35}
    \alpha_{ s_{\alpha} }= \left( s_{\alpha} - \frac{1}{2} \right)\Delta{\alpha},\text{~~}s_{\alpha} = 1,\ldots ,\left \lfloor S_{\alpha}\right \rfloor.
\end{equation}
And the $s_{\beta}$-th sampling points in the $\beta$ domain can be reformulated as
\begin{equation}\label{eq36}
    \beta{ s_{\beta} } = - 1 + \left( {s_{\beta} - \frac{1}{2}} \right)\Delta{\beta},\quad s_{\beta} = 1\ldots,\left \lfloor S_{\beta}\right \rfloor.
\end{equation}


(\ref{eq33}) shows that the number of quantized bits of the codeword is only related to the channel correlation $c$ and the number $N$ of antennas but is independent of the frequency. The number of codewords in the $\alpha$ domain is proportional to $\sqrt{N}$, and the number of codewords in the $\beta$ domain is proportional to $N$. Moreover, if the number of antennas $N$ remains unchanged, an increase in the channel correlation $c$ can result in a greater number of codebook quantization vectors.

\subsection{Dislocation Quantization Codebook Scheme}
\begin{figure}
  \centering
  \subfloat[]{
  \label{fig8}
  \includegraphics[width=1.72in]{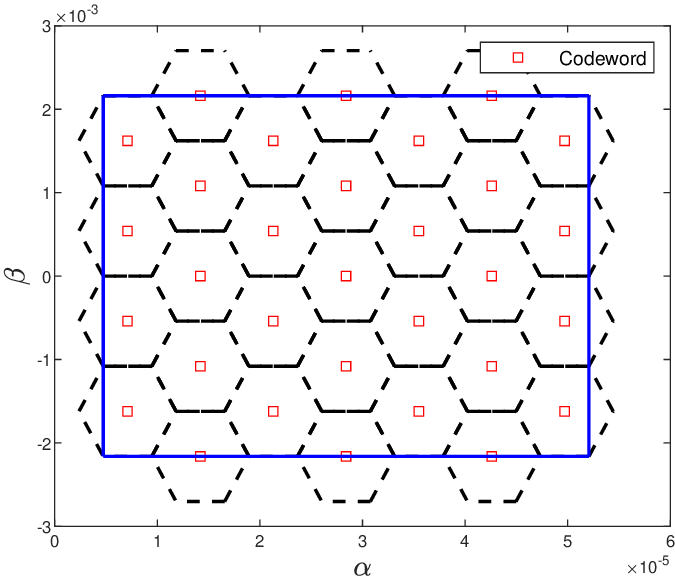}
  }
  \subfloat[]{
  \label{fig9}
  \includegraphics[width=1.72in]{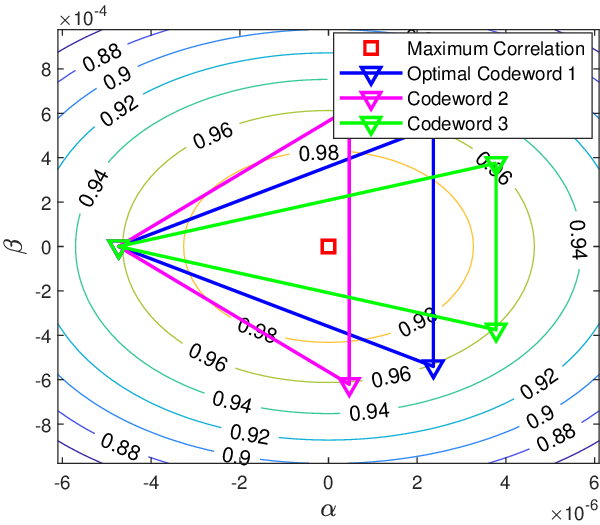}}
  \caption{ULA dislocation codebook: (a) Resource division in the $\alpha$-$\beta$ domain. (b) Quantization area of a codeword with minimum correlation $c$.}
  \label{fig:ENISA2}
\end{figure}

In this section, we propose a dislocation ULA codebook to further improve the quantized accuracy of codewords. The dislocation codebook can be viewed as a combination of two sets of uniform codebooks, as shown in Fig. \ref{fig:ENISA2}\subref{fig8}. $\overline {\Delta \alpha}$ and $\overline{\Delta \beta}$
are the sampling steps of the dislocation codebook in the $\alpha$ and $\beta$ domains. Notably, in comparison to uniform sampling, the dislocated sampling approach introduces a distinct characteristic in the $\alpha$ domain: the $\beta$ value of two adjoining columns of sampling points consistently differs by $\frac{\overline{\Delta \beta}}{2}$. 



The quantization area of each codeword is distributed in a regular hexagon. Three non-adjacent points within the hexagon can form a triangle, as shown in Fig. \ref{fig:ENISA2}\subref{fig9}. The area of the triangle is always half the area of the hexagon. Therefore, the problem of maximizing the quantization area of a dislocation codeword can be transformed into the problem of finding the max inscribed triangle of an ellipse. Consider a vertex $(\alpha^\star,\beta^\star)$ of the triangle in the Fig. \ref{fig:ENISA2}\subref{fig9}, where $\alpha^\star\!>\!0$ and $\beta^\star\!>\!0$.
The area of the inscribed triangle is largest inscribed when the point locates at $\alpha^\star\!=\!\frac{1}{2N^{2}}\!\sqrt{\frac{\left( {c - 1} \right)}{ {p}_{\alpha}}}$ and $\beta^\star\!=\!\frac{1}{2N}\!\sqrt{\frac{3\left( {c - 1} \right)}{{p}_{\beta}}}$. 
Therefore, the optimal sampling steps of dislocation ULA codebook in the $\alpha$ domain and the $\beta$ domain can be calculated as
\begin{equation}\label{eq39}
    \overline{\Delta{\alpha}}\! = \!6\alpha^\star\!=\!\frac{3}{N^{2}}\sqrt{\frac{\left( {c - 1} \right)}{ {p}_{\alpha}}}, \quad
     \overline{\Delta{\beta}}\!=\! 2\beta^\star\!=\! \frac{1}{N}\sqrt{\frac{3\left( {c - 1} \right)}{{p}_{\beta}}}.
\end{equation}
The number of the sampling points in the $\alpha$ and $\beta$ domains is 
\begin{equation}\label{eq41a}
{\overline{S}_{\alpha}} = \frac{1}{3}\sqrt{\frac{Np_{\alpha}}{\left( {c-1} \right)}}~ ,  \quad
{\overline{S}_{\beta}}=2N\sqrt{\frac{p_{\beta } }{3\left ( c-1 \right ) } }.
\end{equation}
Therefore, the total number of sampling points can be calculated as
\begin{equation}\label{eq41b}
\overline{S}_{ULA}=2{\overline{S}_{\alpha}}{\overline{S}_{\beta}}=\frac{4N}{3(1-c)}\sqrt{\frac{Np_\alpha p_\beta}{3}}.
\end{equation}
The $\overline{s}_\alpha$-th sampling point in the $\alpha$ domain is
\begin{equation}\label{eq42}
{\alpha}_{ \overline{s}_\alpha  }=\left \{\begin{matrix}
 {  \frac{2\overline{\Delta\alpha}}{3}+\left ( \overline{s}_\alpha - 1 \right)\overline{\Delta{\alpha}},\quad \overline{s}_\beta \sim odd }& \\
 { \frac{\overline{\Delta\alpha}}{6} +\left( \overline{s}_\alpha - 1 \right)\overline{\Delta{\alpha}}, \quad \overline{s}_\beta \sim even}
\end{matrix} \right.  
\end{equation}
where $\overline{s}_{\alpha} = 1\ldots 
 \left \lfloor\overline{S}_{\alpha}\right \rfloor$. 
And the $s_\beta$-th sampling point in the $\beta$ domain is
\begin{equation}\label{eq43a}
{\beta}_{ \overline{s}_\beta  }=\left \{\begin{matrix}
 {  -1+\left ( \overline{s}_\beta - 1 \right)\overline{\Delta{\beta}},\quad \overline{s}_\alpha \sim odd }& \\
 { -1 +\left( \overline{s}_\beta - \frac{1}{2}\right)\overline{\Delta{\beta}}, \quad \overline{s}_\alpha \sim even}
\end{matrix} \right.  
\end{equation}

(\ref{eq33}) and (\ref{eq41b}) provide the number of codewords under uniform and dislocation sampling. The number of codewords in the dislocation sampling scheme is only $75\%$ of the number of uniform sampling codewords. Therefore, in the same space, the dislocation quantization scheme achieves the goal of low codebook quantization overhead. We compare the number of sampling points in the $\alpha$ and $\beta$ domains of the two schemes in Example \ref{ex1}. 

\begin{example} 
\label{ex1}
\rm
In the case of the same channel correlation $c=0.95$ and different antenna numbers $N$, the optimal codeword numbers for the $\alpha$ and $\beta$ domains are summarized in Table \ref{table1}, respectively. 

It should be noted that, the number of sampling points in the $\alpha$ domain is significantly higher than that in the $\beta$ domain. This phenomenon highlights the robustness of near-field beamforming in the $\beta$ domain, and the denser sampling of the $\beta$ domain enhances codeword quantization accuracy.  Consequently, with the same amount of feedback bits, dense sampling in the $\beta$ domain will be more conducive to improving the quantization performance of the codeword.
\begin{table}[ht]
\setlength\tabcolsep{3pt}
\centering
\caption{Comparison of ULA codebook sampling points.}
\label{table1}
\scalebox{1}{
    \begin{tabular}{|c|c|c|c|c|}
    \hline
        \multirow{2}*{ Number of antenna}& \multicolumn{2}{c|}{Uniform codebook} & \multicolumn{2}{c|}{Dislocation codebook}  \\ 
        \cline{2-5}
        ~ &  $\alpha$ domain & $\beta$ domain& $\alpha$ domain & $\beta$ domain \\ \hline
        64 & 5 & 253& 2 & 412 \\ \hline
        128 & 7 & 507 & 3 & 828 \\ \hline
        256 & 9& 1014 & 4 & 1654\\ \hline
        512 & 13  & 2027 & 5  & 3310  \\ \hline
    \end{tabular}
    }
\end{table}
\end{example}

\section{Near-field UPA Codebook Design}
\label{sec:5}
This section will provide the UPA codebooks with uniform sampling and dislocation sampling, respectively. The non-stationary feature of the UPA channel leads to varying quantization performance for each codeword. To address this issue, we initiate by defining a reference ellipsoid and assume that the correlation formula for each codeword is always the same as the reference ellipsoid. Based on this assumption of stationarity, the uniform codebook and dislocation codebook can be obtained. The performance of the UPA codebook based on the reference ellipsoid is the lower bound of quantization performance under the assumption of stationarity. In other words, the actual quantization area of each codeword is all within the reference ellipsoid of each UPA codebook.


Corollary \ref{Propo1} demonstrates that for any codeword, the pointing position of the channel vector that satisfies the minimum quantization correlation of $c$ is always uniformly distributed on an ellipsoid centered around the quantization center of the codeword.  It's important to note that, due to the non-stationary characteristics of UPA channels, the size of the ellipsoid enclosed by different codewords meeting the same minimum quantization correlation conditions varies. Consequently, when designing the optimal sampling interval between UPA codewords, we can't directly apply the correlation feature of any single codeword to all codewords, as is the case with ULA codebooks. In order to solve the problem caused by non-stationary features when designing the optimal sampling interval for codebooks, we hope to find a reference ellipsoid to describe the quantization features of any codeword in space. This reference ellipsoid provides the maximum allowable space, ensuring that all codewords can guarantee the minimum quantization correlation $c$ at this volume. Below, we define a reference ellipsoid.
\begin{myDef}
\rm
Consider sampling $T_{\psi }$, $T_{\varphi }$ and $T_{\rho }$ points on the $\psi$, $\varphi$ and $\rho$ domains, respectively. From Corollary \ref{Propo2}, it can be concluded that when the quantization performance of all codewords satisfies the minimum correlation $c^{\mathrm{\star}}$, the sets of ellipsoidal axis lengths enclosed by the quantization boundaries of all codewords are respectively represented as $\mathbf{L}_{\psi }\!=\! \left\{l_{\psi ,1},\ldots ,l_{\psi ,{T_{\psi }}}\right\}$, $\mathbf{L}_{\varphi }\!=\!\left\{l_{\varphi ,1},\ldots , l_{{\varphi,T_{\varphi }}}\right\}$, $\mathbf{L}_{\rho }\!=\!\left\{l_{\rho ,1},\ldots , l_{{\rho ,T_{\rho }}}\right\}$. Meanwhile, ${l}_{\psi }^{\mathrm{\star}}\!=\! \mathit{\min } \ \mathbf{L}_{\psi }$, ${l}_{\varphi }^{\mathrm{\star}}\!=\!\mathit{\min } \ \mathbf{L}_{\varphi }$ and ${l}_{\rho }^{\mathrm{\star}}\!=\! \mathit{\min } \ \mathbf{L}_{\rho }$ are used as the axial length of the reference ellipsoid. Thus, the formula of the reference ellipsoid can be written as
\begin{equation}\label{upaep}
\frac{\delta_{\psi_{s}}^{2}}{\big({l}_{\psi }^{\mathrm{\star}}\big)^{2}}+\frac{\delta_{\varphi_{s}}^{2}}{\big({l}_{\varphi }^{\mathrm{\star}}\big)^{2}}+\frac{\delta_{\rho_{s}}^{2}}{\big({l}_{\rho }^{\mathrm{\star}}\big)^{2}}=1.
\end{equation}
According to the fitting formula given in Corollary \ref{Propo2}, the fitting coefficient can be calculated as 
\begin{equation}
\begin{split}
    {p}_{\psi }^{\star}=\frac{\mathrm{c}-1}{({l}_{\psi }^{\star}N)^{2}},\ \quad
    {p}_{\varphi }^{\star}=\frac{\mathrm{c}-1}{({l}_{\varphi }^{\star}N)^{2}}, \ \quad
    {p}_{\rho}^{\star}=\frac{\mathrm{c}-1}{({l}_{\rho}^{\star}N^2)^{2}}.
\end{split}
\end{equation}
The formula for the reference ellipsoid can be completed as
\begin{equation}\label{upaep1}
\frac{\delta_{\psi_{s}}^{2}}{\frac{\mathrm{c}-1}{{p}_{\psi }^{\mathrm{\star}}N^{2}}}+\frac{\delta_{\varphi_{s}}^{2}}{\frac{\mathrm{c}-1}{{p}_{\varphi }^{\mathrm{\star}}N^{2}}}+\frac{\delta_{\rho_{s}}^{2}}{\frac{\mathrm{c}-1}{{p}_{\rho }^{\mathrm{\star}}N^{4}}}=1.
\end{equation}
\end{myDef}


It is important to note that the reference ellipsoid is a virtual area reconstructed by considering the minimum value of all codeword quantization areas. Therefore, codewords with the reference ellipsoid as the quantization area may not exist. The minimum correlation of certain codewords may be smaller than $c$ if the reference ellipsoid becomes larger. Thus, the reference ellipsoid represents the largest shape that can describe the quantization areas of all codewords in the UPA channel. Smaller reference ellipsoids allow for a higher minimum correlation for each codeword. Since the reference ellipsoid represents the largest ellipsoid achievable under the assumption of stationarity, it sets a lower bound on the performance achievable by codebook schemes based on the assumption of stationarity.

\begin{figure}
  \centering
  \subfloat[]{
  \label{fig888}
  \includegraphics[width=1.72in]{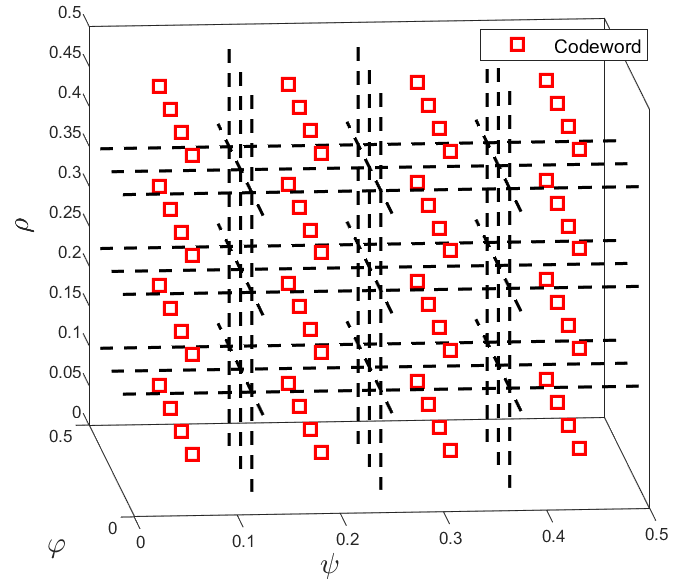}
  }
  \subfloat[]{
  \label{fig999}
  \includegraphics[width=1.72in]{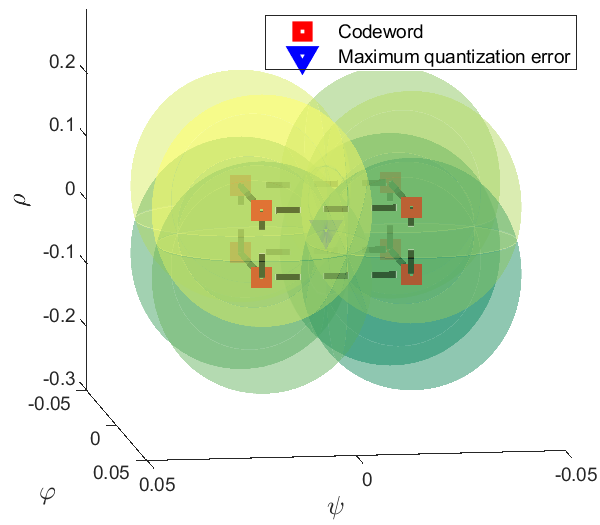}}
  \caption{UPA dislocation codebook: (a) Resource division in the $\varphi$-$\psi$-$\rho$ domain. (b) Quantization area of a codeword with minimum correlation $c$.}
  \label{fig:ENISA3}
\end{figure}

\subsection{Uniform Codebook Quantization Scheme }
In this section, we propose the UPA uniform codebook scheme. The codebook scheme is uniformly sampled in the $\psi$, $\varphi$ and $\rho$ domains, as shown in Fig. \ref{fig:ENISA3}\subref{fig888}. Among them, the red rectangle represents the codeword. The actual quantization area of each codeword is a cuboid. Under the assumption of a stationary UPA channel, the quantization areas of codewords can be represented using a reference ellipsoid expressed by (\ref{upaep1}). The quantization area of a codeword is the inscribed cuboid of the ellipsoid. For adjacent eight codewords, the boundaries of their quantization regions consistently intersect at a single point, as shown in Fig. \ref{fig:ENISA3}\subref{fig999}. Therefore, the problem of maximizing codeword quantization regions can be transformed into finding the maximum inscribed rectangular cuboid of an ellipsoid. 

Taking codeword $\check{\mathbf{w}}\!=\!\mathbf{b}^*(0,0,0)$ as an example, $(\psi^\star,\!\varphi^\star,\! \rho^\star)$ represents a vertex located on the reference ellipsoid corresponding to this codeword. Here, it's crucial to note that $\psi^\star\!>\!0$, $\varphi^\star\!>\!0$, and $\rho^\star\!>\!0$. Under the conditions $\psi^\star\!=\!\frac{1}{N}\sqrt{\frac{\mathrm{c}-1}{3{p}_{\psi }^{\mathrm{\star}}}}$, $\varphi^\star\!=\!\frac{1}{N}\sqrt{\frac{\mathrm{c}-1}{3{p}_{\varphi }^{\mathrm{\star}}}}$ and $\rho^\star\!=\!\frac{1}{N}\sqrt{\frac{\mathrm{c}-1}{3{p}_{\rho }^{\mathrm{\star}}}}$, the volume of the inscribed rectangular cuboid within the ellipsoid reaches its maximum. Consequently, the quantization space of the codeword achieves its maximum extent.
In such a scenario, the optimal sampling steps can be calculated as
\begin{equation}
\begin{split}
\Delta{\psi }=2\psi^\star=\frac{2\sqrt{3}}{3N}\sqrt{\frac{\mathrm{c}-1}{{p}_{\psi }^{\mathrm{\star}}}},\\
\Delta{\varphi }=2\varphi^\star=\frac{2\sqrt{3}}{3N}\sqrt{\frac{\mathrm{c}-1}{{p}_{\varphi }^{\mathrm{\star}}}}, \\
\Delta{\rho }=2\rho^\star=\frac{2\sqrt{3}}{3N^{2}}\sqrt{\frac{\mathrm{c}-1}{{p}_{\rho }^{\mathrm{\star}}}}.
\end{split}
\end{equation}
The codebook is designed for a 3D space with a distance range of $r\in\big[ 0.62\sqrt{\frac{D^{3}}{\lambda} }  ,\infty  \big )$, elevation angle of $\left[-\frac{\pi }{2},\frac{\pi }{2}\right]$ and azimuth angle of $\left[0,\pi \right]$. And the range of the 3D space in the transformed domain is given by 
\begin{equation}\label{rangg}
Q_\psi=2,\quad Q_\varphi=2,\quad Q_\rho\approx \frac{2.7}{N\sqrt{N}}.
\end{equation}
Therefore, the number of codewords in UPA uniform codebook can be calculated as
\begin{equation}
S_{\psi }=\sqrt{\frac{3{p}_{\psi }^{\mathrm{\star}}}{c-1}}N ,  S_{\varphi }= \sqrt{\frac{3{p}_{\varphi }^{\mathrm{\star}}}{c-1}}N ,  S_{\rho }\! \approx 2.3\sqrt{\frac{N{p}_{\rho }^{\mathrm{\star}}}{c-1}}.
\end{equation}
The positions represented by the $s_{\psi }$-th, $s_{\varphi }$-th and $s_{\rho }$-th sampling points in $\psi $, $\varphi$ and $\rho$ domain can be expressed as
\begin{equation}
\begin{aligned}
\psi_{s_{\psi }} & = -1 + \left(s_{\psi } - \frac{1}{2}\right) \Delta{\psi }, \quad s_{\psi } = 1, \ldots ,\left \lfloor S_{\psi } \right \rfloor , \\
\varphi_{s_{\varphi }} & = -1 + \left(s_{\varphi } - \frac{1}{2}\right) \Delta{\varphi }, \quad s_{\varphi } = 1, \ldots ,\left \lfloor S_{\varphi }\right \rfloor , \\
\rho_{ s_{\rho }} & = \left(s_{\rho } - \frac{1}{2}\right) \Delta{\rho }, \quad s_{\rho } = 1, \ldots ,\left \lfloor S_{\rho }\right \rfloor .
\end{aligned}
\end{equation}
And the total number of sampling points is  
\begin{equation}
\begin{split}
S_{max} =  \frac{7N^2}{(1-c)}\sqrt{\frac{N{p}_{\psi }^{\mathrm{\star}}{p}_{\varphi }^{\mathrm{\star}}{p}_{\rho }^{\mathrm{\star}}}{c-1}}.
\end{split}
\end{equation}

For the proposed uniform codebook scheme, the number of sampling points is proportional to the number of antennas and minimum quantization correlation. And the sampling step will decrease with the increased minimum quantization correlation and the number of antennas.

\subsection{Dislocation Quantization Codebook Scheme }
\begin{figure}
  \centering
  \subfloat[]{
  \label{fig81}
  \includegraphics[width=1.72in]{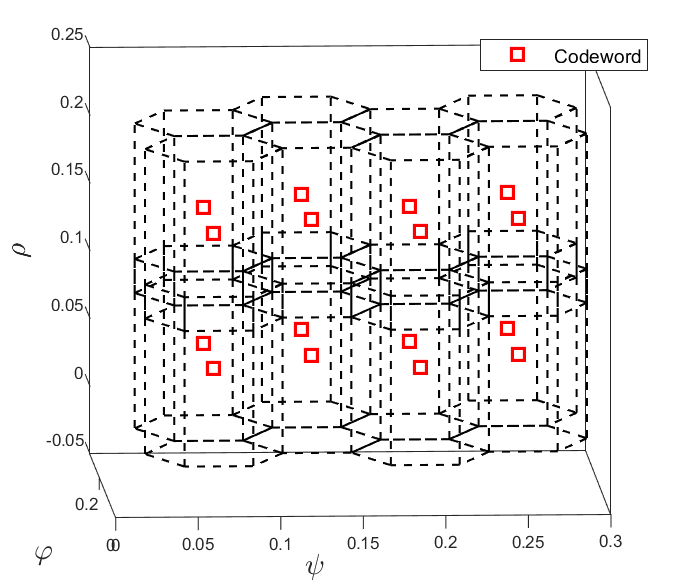}
  }
  \subfloat[]{
  \label{fig91}
  \includegraphics[width=1.72in]{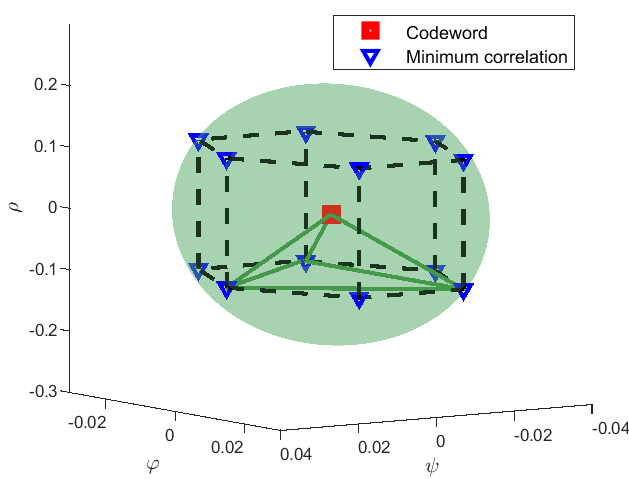}}
  \caption{UPA uniform codebook: (a) Resource division in the $\varphi$-$\psi$-$\rho$ domain. (b) Quantization area of a codeword with minimum correlation $c$.}
  \label{fig:ENISA4}
\end{figure}

In this section, we will explore a UPA dislocation codebook to further decrease the quantization overhead. The quantization area of UPA dislocation codeword is a hexagonal prism, illustrated in the Fig. \ref{fig:ENISA4}\subref{fig81}. The UPA dislocation codebook can be viewed as a combination of two same sets of UPA uniform codebooks. Here, $\overline{\Delta{\psi}}$, $\overline{\Delta{\varphi}}$, and $\overline{\Delta{\rho}}$ denote the sampling intervals of the UPA dislocation codebook in the $\psi$, $\varphi$, and $\rho$ domains, respectively. By shifting adjacent sampling points in the $\psi$ domain of a uniform codebook by $\delta/2$ in the $\varphi$ domain, a dislocated UPA codebook can be obtained. 


Fig. \ref{fig:ENISA4}\subref{fig91} is the quantization performance of the codeword $\check{\mathbf{w}}$ considering a minimum correlation of $c$. The optimization of the quantization area for the UPA dislocated codebook can be reformulated as the task of maximizing the volume of an inscribed hexagonal prism within the quantization area. We select the inscribed hexagonal prism in the Fig. \ref{fig:ENISA4}\subref{fig91}. A vertex, denoted as $(\psi^\star,\varphi^\star,\rho^\star)$, is a vertex located on the triangular pyramid satisfying conditions $\rho^\star\!<\!0$, $\psi^\star\!>\!0$ and $\varphi^\star\!>\!0$. The quantization space reaches its maximum extent when $\rho^\star\!=\!\sqrt{\frac{c-1}{3p_\rho^\star N^4}}$ and $\psi^\star\!=\!\sqrt{\frac{c-1}{6p_\psi^\star N^2}} $, and $\varphi^\star\!=\!\sqrt{\frac{c-1}{2p_\varphi^\star N^2}} $. Thus, the optimal quantization intervals can be calculated as
\begin{equation}
\begin{split}
\overline{\Delta{\psi }}=\frac{1}{N}\sqrt{\frac{6(\mathrm{c}-1)}{{p}_{\psi }^{\mathrm{\star}}}},\\
\overline{\Delta{\varphi }}=\frac{1}{N}\sqrt{\frac{2(c-1)
}{{p}_{\varphi }^{\mathrm{\star}}}}, \\
\overline{\Delta{\rho }}=\frac{2\sqrt{3}}{3N^{2}}\sqrt{\frac{c-1}{{p}_{\rho }^{\mathrm{\star}}}}.
\end{split}
\end{equation}
The number of sampling points in the $\psi$, $\varphi$, and $\rho$ domains are respectively
\begin{equation}
\overline S_{\psi }= 2\sqrt{\frac{{p}_{\psi }^{\mathrm{\star}}}{6(c-1)}}N,  \overline 
S_{\varphi }= \sqrt{\frac{2{p}_{\varphi }^{\mathrm{\star}}}{c-1}}N, \overline 
S_{\rho }\approx 2.3\sqrt{\frac{N{p}_{\rho }^{\mathrm{\star}}}{c-1}}.
\end{equation}
The $\overline{s}_{\rho }$-th sampling point in the $\rho$ domain is expressed as
\begin{equation}
\begin{split}
\rho _{ \overline s_{\rho }}  &=\left(\overline s_{\rho }-\frac{1} {2}\right)\Delta{\rho },\quad \overline s_{\rho }=1,\ldots ,\left \lfloor \overline S_{\rho }\right \rfloor.
\end{split}
\end{equation}
The $\overline{s}_{\varphi }$-th sampling point in the $\varphi$ domain is calculated as 
\begin{equation}\label{eq43aba}
 {\psi}_{\overline{s}_{\varphi } } =\left \{\begin{matrix}
&{- 1 +  \left( {\overline{s}_{\varphi }- 1} \right)\overline{\Delta{\varphi }},\quad \overline s_{\psi }\sim even} \\
& {- 1 + \frac{\overline{\Delta\varphi }}{2} + \left( {\overline{s}_{\varphi } - 1} \right)\overline{\Delta{\varphi}}},\quad \overline s_{\psi }\sim odd
\end{matrix}\right . ,
\end{equation}
And the $\overline{s}_{\psi }$-th sampling point in the $\psi$ domain is 
\begin{equation}\label{eq43ac}
 {\psi}_{\overline{s}_{\psi } } =\left \{\begin{matrix}
&{- 1 +  \left( {\overline{s}_{\psi }- 1} \right)\overline{\Delta{\psi }},\quad \overline s_{\varphi }\sim odd} \\
& {- 1 + \frac{\overline{\Delta\psi }}{2} + \left( {\overline{s}_{\psi } - 1} \right)\overline{\Delta{\psi}}},\quad \overline s_{\varphi }\sim even
\end{matrix}\right . ,
\end{equation}
where $\overline{s}_\psi = 1\ldots,\left \lfloor {\overline{S}}_{\psi} \right \rfloor  $.
Therefore, the total number of sampling points of dislocated codewords is calculated as
\begin{equation}
\begin{split}
\overline S_{max} = \frac{5.3N^2}{(1-c)}\sqrt{\frac{N{p}_{\psi }^{\mathrm{\star}}{p}_{\varphi }^{\mathrm{\star}}{p}_{\rho }^{\mathrm{\star}}}{c-1}} . 
\end{split}
\end{equation}
It can be seen that $\frac{\overline S_{UPA}}{S_{UPA}}\approx 0.75$. Therefore, the overhead of dislocation codebook is always only 0.75 times that of uniform codewords under the same quantization area. This advantageous aspect is highlighted through Example \ref{ex2}, where we conduct a comparison of the sampling points between the proposed UPA codebook schemes.
\begin{example}\label{ex2}
\rm
Table \ref{table2} illustrates the number of sampling points for the proposed UPA codebook schemes with minimum correlation $c =0.95$ and frequency $f=100$ GHz. 
Without loss of generality, the quantization overhead of UPA dislocation codebooks is always smaller than UPA uniform codebooks in the same system configuration. Of noteworthy significance is the consistent trend where the number of codewords in the $\psi$ and $\varphi$ domains consistently surpasses those in the $\rho$ domain. This underscores that the robustness of the angle domain is stronger than the distance domain in the UPA channel. 
\begin{table}[htbp]
\setlength\tabcolsep{3pt}
\centering
\caption{{Comparison of UPA codebook sampling points.}}
\label{table2}
\scalebox{1}{
    \begin{tabular}{|c|c|c|c|c|}
    \hline
        \multirow{2}*{ Number of antenna}& \multicolumn{2}{c|}{Uniform codebook} & \multicolumn{2}{c|}{Dislocation codebook}  \\ 
        \cline{2-5}
        ~ &  $\psi/\varphi$ domain & $\rho$ domain  & $\psi/\varphi $ domain & $\rho$ domain \\ \hline
        8*8 & 39*39 & 2 & 18*32 & 2 \\ \hline
        12*12 & 58*58 & 3 & 27*47 & 3 \\ \hline
        16*16 & 78*78 & 3 & 37*64 & 3\\ \hline
    \end{tabular}
    }
\end{table}
\end{example}

\section{SIMULATION RESULTS }
\label{sec:6}
In this section, we provide the simulation results to illustrate the performance of the proposed codebook schemes for ULA and UPA systems. The simulation considers the ULA system in Fig. \ref{fig1}. In the ULA system, the number of the transmitted antenna is set as $N_1=512$. The UE locates randomly in the space spanned $\big ( r_1,\theta_1  \big )  \in\big [ 0.62\sqrt{\frac{(N_1d)^{3}}{\lambda }}  ,\infty  \big ) \times \big [ -\frac{\pi}{2},\frac{\pi} {2} \big ]$.  
And the UPA system in Fig. \ref{fig22} is used for simulation. In the UPA system, the number of the transmitted antenna is $N_2\times N_2$ with $N_2=16$. The elevation angle and azimuth angle of the UE is $\theta_2 \in \left[-\frac{\pi }{2},\frac{\pi }{2}\right]$ and $\phi_2\in\left [0,{\pi }\right]$, respectively. 
The distance between the BS and the UE distributes in $r_2\in\big [ 0.62\sqrt{\frac{D_2^{3}}{\lambda} },\infty  \big )$. The carrier frequency of both ULA and UPA systems is set to $f=\SI{100}{GHz}$. UE may be in the near-field or far-field with the above configuration. 

We evaluate the cumulative probability function (CDF) and achievable rate of the proposed codebook schemes. 
The signal-to-noise ratios (SNR) of the ELAA system can be calculated as \cite{ref7}
\begin{equation}
\mathrm{SNR}=\frac{P\eta N}{r^{2}\sigma ^{2}},
\end{equation}
where $P$ is the transmit power and $\sigma^2$ is the noise power set as $\sigma^2=\SI{-70}{dBm} $. The achievable rate is given by 
\begin{equation}
R=\log_{2}\left ( 1+\frac{P\eta N \left | \mathbf b^T\left ( r,\theta  \right ) \mathbf w\right | ^{2} }{r^{2}\sigma  ^{2}  }  \right ).
\end{equation}
And the simulation results are the average results of 1000 randomly distributed UE. 

\subsection{ULA Codebook }
\begin{figure}[t]
\centering
\includegraphics[width=3.5in]{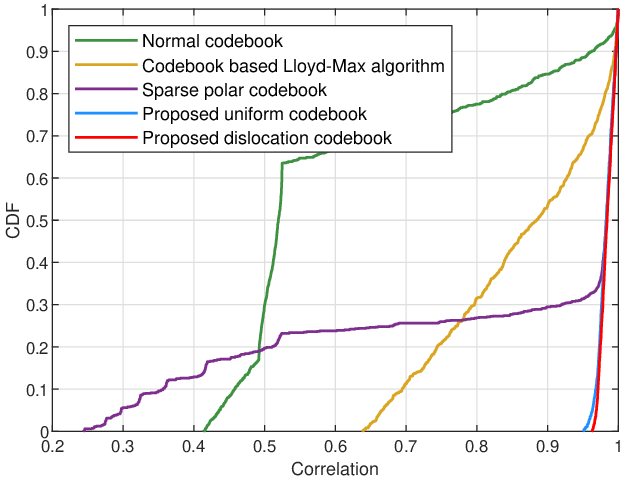}
\caption{CDF of the codebook quantification correlation in ULA channel, with $N_1 = 512$ and $f = \SI{100}{GHz}$.\label{Fig.12}}
\end{figure}

Fig. \ref{Fig.12} illustrates the CDF of quantized correlation with various codebook schemes. In order to provide a comparative analysis, the proposed codebook schemes are compared with the following schemes:
\begin{itemize}
    \item Normal codebook: The codebook uniformly samples the $\alpha$ and $\beta$ domains. And the number of sampling points in the $\alpha$ and $\beta$ domains are both $\frac{N_1}{3}$.   
    \item Codebook based on Lloyd-Max algorithm: 
    The Lloyd-Max algorithm is used to sample $\alpha$ and $\beta$ domains\cite{ref19}. The number of sampling points for $\alpha$ domain is designed as $\frac{N_1}{10}$; for $\beta$ domain, it is set as $N_1$.
    \item Sparse polar codebook: The codebook is sampled on a sparse domain of distance domain and angle domain \cite{ref20}. The number of the codewords is $S_P=\sum_{n=1}^{N_1} S^{(n)}_P$, where $N_1$ is the sampling number in the $\alpha$ domain and $S^{(n)}_P$ is the sampling number in the $\beta$ domain. $S_P$ can be calculated as $28431$ in the considered ULA system.
\end{itemize}
For equality, the proposed ULA dislocation codebook and uniform schemes are quantized with $14.8$ bits. The proposed uniform codebook comprised 2027 sampling points in the $\alpha$ domain and 14 sampling points in the $\beta$ domain. On the other hand, the dislocation quantization codeword contains 1111 sampling points in the $\alpha$ domain and 13 sampling points in the $\beta$ domain. It is worth noting that the dislocation codebook consistently outperforms the uniform codebook under the equal number of codewords. The performance of the ULA normal codebook is considerably inferior to the proposed codebook schemes, even when employing a more significant number of quantization vectors. Furthermore, the proposed schemes demonstrate significant superiority over the codebook based Lloyd-Max algorithm and sparse polar domain codebook, which validates the effectiveness of our proposed schemes. In addition, the minimum quantization correlation of the proposed codebook in \cite{ref20} is smaller than other codebook schemes. Moreover, our simulation includes an infinite distance, which verifies that the proposed solution also has good applicability in far-field scenarios.

\begin{figure}[t]
\centering
\includegraphics[width=3.5in]{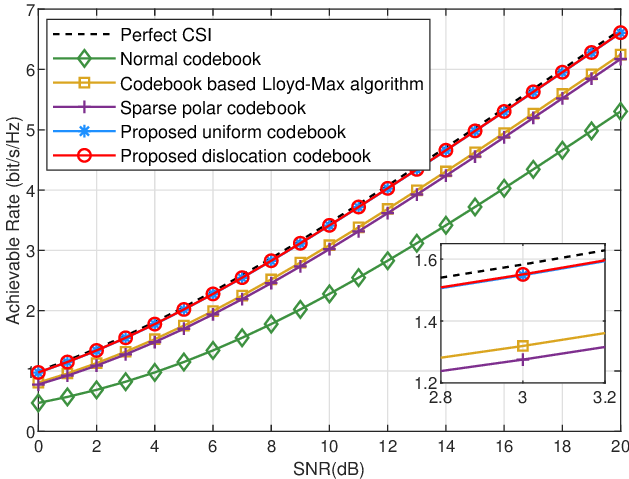}
\caption{Achievable rate against SNR for several codebook schemes in ULA channel, with $N _1= 512$ and $f = \SI{100}{GHz}$.\label{Fig.13}}
\end{figure}
Fig. \ref{Fig.13} illustrates the achievable rate for two scenarios: the ideal case of perfect CSI and the case that the precoding matrix is selected based on codebook. The beamforming scheme with perfect CSI represents the theoretical upper limit. In this comparison, we consider the same codebook schemes shown in Fig. \ref{Fig.12}. The proposed codebooks and the ideal case of perfect CSI exhibit remarkably similar performance. Notably, the dislocation codebook significantly enhances the achievable rate compared to the uniform codebook. When employing the same quantization bits, we observe that the achievable rate of the proposed codebooks consistently outperforms the rate achieved by other codebook schemes, particularly as the receiver SNR increases. With $\mathrm{SNR}=20$ dB, the achievable rate demonstrates an improvement of approximately $1.4$ bit/s/Hz compared to the ULA normal codebook.

\subsection{UPA Codebook }
\begin{figure}[t]
\centering
\includegraphics[width=3.5in]{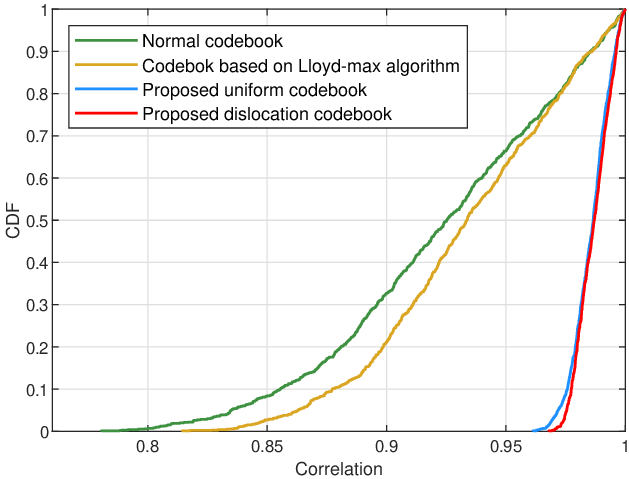}
\caption{CDF of the codeword quantification correlation in UPA channel, with $N= 16\times 16$, $f=\SI{100}{GHz}$.\label{Fig.15}}
\end{figure}

Fig. \ref{Fig.15} shows the CDF of the quantification correlation of the UPA codebook schemes. We compare the proposed UPA codebook schemes with the other two schemes:
\begin{itemize}
    \item  Normal codebook: This codebook employs identical sampling points in the $\psi$, $\varphi$, and $\rho$ domains, consisting of 29 uniform samples within each of these domains.
    \item  Codebook based Lloyd-Max algorithm: The angle domain is sampled as the traditional far-field codebook, and the distance domain is sampled using the Lloyd-Max algorithm. The number of sampling points in the $\psi$, $\varphi$, and $\rho$ domains is $N_2$, $N_2$, and $3N_2$, respectively \cite{ref19}.
\end{itemize}
The minimum quantization correlation of the uniform and codebook schemes are set as $0.95$ and $0.96$, respectively. The quantization precision of the UPA uniform scheme and UPA dislocation codebook scheme is approximately of 14 bits and a half. The simulation results consistently demonstrate that the quantized correlation achieved by the proposed codebook consistently exceeds the predefined minimum correlation.  The results indicate that the quantization performance of UPA dislocation codebook schemes is better than that of uniform codebook schemes while maintaining the same quantization overhead. Our proposed UPA codebook schemes achieve superior quantization correlation compared to the codebook designed using the Lloyd-Max algorithm. Moreover, the results also reveal that the performance of our proposed codebook consistently outperforms the UPA normal scheme which sets as equal number of samples in the $\psi$, $\varphi$, and $\rho$ domains. The result shows that dense sampling in the angle domain is more conducive to accurate quantization of UPA channels. On the contrary, the distance domain only requires a small number of bits for quantization. 

\begin{figure}[t]
\centering

\includegraphics[width=3.5in]{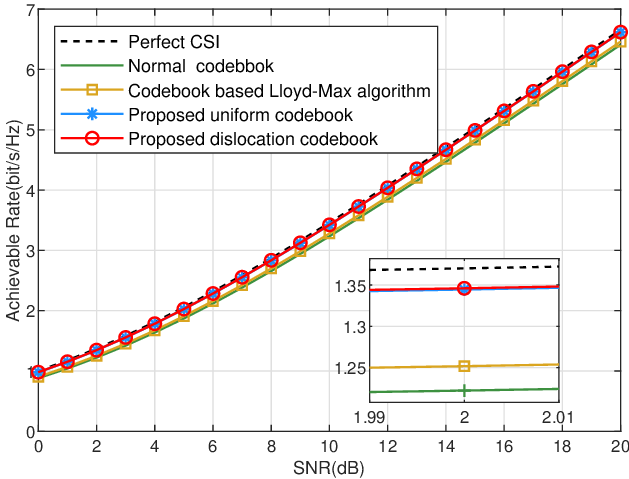}
\caption{Achievable rate against SNR for several codebooks in UPA channel, with $N= 16\times 16$, $f=\SI{100}{GHz}$.\label{Fig.16}}
\end{figure}

We use different codebook schemes as the beamforming matrix at the BS in the UPA system and evaluate the achievable rate. Fig. \ref{Fig.16} illustrates the results with various SNR. The outcomes demonstrate that our proposed UPA codebook schemes outperform the other two schemes and closely approach the performance achieved with perfect CSI. At an SNR of $20$ dB, our proposed schemes exhibit an enhancement of approximately $0.2 $ bit/s/Hz when compared to the UPA normal codebook scheme. 


\section{Conclusion}
\label{sec:7}
This paper introduces a novel codebook design to maximize the minimum quantization correlation for near-field ELAA channels. 
The performance of the ULA codebook is stationary and symmetrical. Moreover, the performance of the UPA codebook is non-stationary and asymmetrical. The correlation formula of ULA and UPA can be fitting as ellipse and ellipsoid, respectively. 
Based on these insights, we propose two ULA codebooks: uniform sampling and dislocation sampling. The dislocation codebook scheme performs like the uniform codebook but needs fewer quantization bits. To address the non-stationarity of the UPA codeword, we propose UPA uniform and dislocation codebook schemes based on the assumption of stationarity. This way, the achievable minimum quantization correlation of the proposed codebook schemes is always greater than that achieved by a single codeword. Additionally, we emphasize the robustness of the angle domain in ELAA systems. Simulation results confirm that the proposed codebook achieves minimal quantization bits while maintaining high quantization performance.

\begin{appendices}

\section{proof of property 1}
\label{proof1}
With the same displacement difference, the correlation formula of $\mathbf{w}_s$ and $\mathbf{w}_{\overline{s}}$ can be calculated as 
\begin{equation}
\begin{split}
f\left(\right. & \left. \alpha_s,\beta_s; \alpha_s+\delta _{\alpha } ,\beta_s+\delta _{\beta } \right)\\
&= \frac{1}{N}\Bigg|\sum_{n=1}^{N}\exp\Bigg( -j\pi \bigg ( \delta _\alpha  n^2 +\Big( \delta _\beta
-\delta _\alpha\big( N+1 \big) \Big) n \bigg)\Bigg) \Bigg|\\
&=f\left(\right.  \left. \alpha_{s^{'}},\beta_{s^{'}}; \alpha_{s^{'}}+\delta _{\alpha } ,\beta_{s^{'}}+\delta _{\beta } \right).
\end{split}
\end{equation} 
Then Property \ref{P11} is proved.
\section{ proof of property 2 }
\label{proof2}
(\ref{eq12}) can be rewritten as
\begin{equation}\label{eq16}
    f\left(\delta _{\alpha } ,\delta _{\beta } \right)=\frac{1}{N}\left|\sum_{n=1}^{N}\exp\left(-j\frac{2\pi }{\lambda }\left(\delta _{\beta }  y_n-\frac{2}{\lambda }\delta _{\alpha }  y_n^2\right)\right) \right|.
\end{equation}
If {${\beta }_q\!-\!{\beta }_s\!=\!-\delta _{\beta } $}, the above formula can be calculated as
\begin{equation}\label{eq17}
    f\left(\delta _{\alpha }  ,-\delta _{\beta }  \right)=\frac{1}{N}\left | \sum_{n=1}^{N}\exp\left(-j\frac{2\pi }{\lambda }\left(-\delta _{\beta }  y_n-\frac{2}{\lambda }\delta _{\alpha }  y_n^2\right)\right)\right |. 
\end{equation}
Since {$y_{N-n+1}=-y_n$}, (\ref{eq17}) can be written as
\begin{equation}
\begin{split}
&f\left(\delta _{\alpha }  ,
-\delta _{\beta }  \right)\\
&=\frac{1}{N}\left|\sum_{n=1}^{N} \exp\left(-j\frac{2\pi }{\lambda }\left(\delta _{\beta }  y_{N-n+1}-\frac{2}{\lambda }\delta _{\alpha }  y_{N-n+1}^2\right)\right) \right|,
\end{split}
\end{equation} 
which indicates that $f\left(\delta _{\alpha }  ,\delta _{\beta } \right) =f\left(\delta _{\alpha }  ,
-\delta _{\beta }  \right)$.
It is evident that {$f(\delta _{\alpha } ,\delta _{\beta } )\!=\!f(-\delta _{\alpha } ,-\delta _{\beta } )$} by central symmetry. Therefore, we can deduce that {$f(\delta _{\alpha } ,\delta _{\beta } )\!=\!f(-\delta _{\alpha } ,\delta _{\beta } )$}. The proof of  Property \ref{pro1} is thus completed.
\section{proof of property 3}
\label{proof3}
According to the (\ref{eq23a}), the quantization performance of codeword always is related to the quantization center $(\psi_s,\varphi_s,\rho_s)$. With the same displacements $(\delta_{\psi_s}, \delta_{\varphi_s},\delta_{\rho_s})$, the quantization performance of tow codewords respectively pointing to $(\psi_s,\varphi_s,\rho_s)$ and $(\psi_s,\varphi_s,\rho_s)$ is different, that is, $f(\psi_s, \varphi_s, \rho_s;\delta_{\psi}, \delta_{\varphi}, \delta_{\rho})\ne f(\psi_{s^{'}}, \varphi_{s^{'}}, \rho_{s^{'}};\delta_{\psi}, \delta_{\varphi}, \delta_{\rho}
)$. In this way, Property \ref{pro3} is proved.

\section{proof of property 4}
\label{proof4}
Replace $\rho_s$ with $\rho_{s^{'}}$ that meets $\rho_q - \rho_{s^{'}} = -\delta_{\rho_s}$. Then, the phase of the $(mN+n)$-th exponential term in (\ref{eq23a}) can be calculated as
\begin{equation}\label{eq26a0}
\begin{split}
\Upsilon _{(m,n)}&(\psi_s, \varphi_s, \rho_s;\delta_{\psi_s},\delta_{\varphi_s},-\delta_{\rho_s}) \\
&=\Upsilon _{(m,n)}(\psi_s, \varphi_s, \rho_s;\delta_{\psi_s},\delta_{\varphi_s},\delta_{\rho_s})+\iota ^{(\rho)}_{(m,n)}
\end{split},
\end{equation}
where $\iota ^{(\rho)}_{(m,n)}$ can be calculated as
\begin{equation}\label{eq26a1}
\begin{split}
\iota ^{(\rho)}_{(m,n)}=&\frac{4\delta_{\rho_s}\bigg(1-\psi_s+\delta_{\psi_s}\bigg)}{\lambda^2} x_m^2+\frac{2\delta_{\rho_s}\bigg(1-\varphi_s+\delta_{\varphi_s}\bigg)}{\lambda^2}y_n^2 \\
&- \delta_{\rho_s}\frac{2\bigg(\psi_s-\delta_{\psi_s}\bigg)\bigg(\varphi_s-\delta_{\varphi_s}\bigg)}{\lambda^2} x_m y_n.
\end{split}
\end{equation}
Therefore, we can obtain that $f\Big(\psi_s,\varphi_s,\rho_s;\!\delta_{\psi_s},\!\delta_{\varphi_s},\!\delta_{\rho_s}\Big)\ne\!f\left(\psi_{s},\varphi_{s},\rho_{s}; \delta_{\psi_s}, \delta_{\varphi_s},- \delta_{\rho_s}\right)$. Using the same method, Property \ref{pro4} can be proved.

\end{appendices}

\end{document}